\documentclass[10pt,aps,prc,floatfix,
twocolumn,superscriptaddress,nofootinbib, preprintnumbers]{revtex4-2}

\usepackage[dvipsnames]{xcolor}
\usepackage{amsfonts,amsmath,amssymb,bm}
\allowdisplaybreaks
\usepackage{graphicx}
\usepackage{microtype}
\usepackage{placeins}
\usepackage{braket}
\usepackage{array}
\usepackage{dcolumn}
\newcolumntype{P}[1]{>{\raggedright\arraybackslash}p{#1}}
\usepackage{cellspace}
\usepackage{xspace}
\usepackage{isotope}
\usepackage{xparse}
\usepackage{physics}
\usepackage[normalem]{ulem}
\usepackage{multirow}
\usepackage{booktabs}
\usepackage{enumitem}
\usepackage[pdfpagelabels, pdfencoding=auto, psdextra]{hyperref}
\hypersetup{%
 pdfsubject=Paper,
 pdfkeywords={nuclear physics, nuclear theory, gpdiff, gaussian process, equation of state, neutron stars, chiral EFT, symmetry energy, incompressibility, FRIB, many-body perturbation theory, MBPT, many-body theory},
 unicode = true,
 breaklinks = true,
 colorlinks = true,
 linkcolor = blue,
 menucolor = blue,
 citecolor = blue,
 urlcolor = blue
}

\definecolor{bobcatgreen}{rgb}{0.3,0.6,0.1}
\definecolor{scarlet}{rgb}{0.73,0,0}

\graphicspath{{./figures/}}

\newcommand{\orcid}[1]{\href{https://orcid.org/#1}{\includegraphics[scale=0.055]{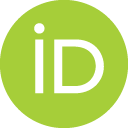}}}
\usepackage{xifthen}

\newcommand{\iso}[2]{#1_#2}

\newcommand{\kF}[1][]{%
  \ifthenelse{\isempty{#1}}%
    {k_\text{F}}
    {k_\text{F}^{(#1)}}
}

\newcommand{\nk}[1][]{%
  \ifthenelse{\isempty{#1}}%
    {n_{k}^{(\tau)}}
    {n_{k_#1}^{(\tau_#1)}}
}

\newcommand{\eps}[1][]{%
  \ifthenelse{\isempty{#1}}%
    {\iso{\varepsilon}{{\tau}}(k)}
    {\iso{\varepsilon}{{\tau_#1}}(k_#1)}
}

\newcommand{\fmiq}{\, \text{fm}^{-3}}

\newcommand{\MeV}{\, \text{MeV}}

\newcommand{\gpdiff}{\texttt{GPDiff}\xspace}
\newcommand{\ncc}{n_{cc}}
\newcommand{\dcc}{\delta_{cc}}
\newcommand{\xcc}{x_{cc}}
\newcommand{\diso}{\delta_{\mathrm{iso}}}

\newcommand{\NNLO}{\ensuremath{{\rm N}{}^2{\rm LO}}\xspace}
\newcommand{\NNNLO}{\ensuremath{{\rm N}{}^3{\rm LO}}\xspace}

\newcommand{\normal}{\mathcal{N}}
\DeclareMathOperator{\pr}{pr}
\newcommand{\given}{\,|\,}  

\DeclareMathOperator*{\cov}{cov}
\DeclareMathOperator*{\diag}{diag}

\begin{document}

\title{A Gaussian Process framework for constraining the nuclear equation of state\texorpdfstring{\\}{} from microscopic calculations with correlated uncertainties}

\author{Y.~G.~Lee~\orcid{0000-0002-4931-4444}}
\email{yl518521@ohio.edu}
\affiliation{Department of Physics and Astronomy, \href{https://ror.org/01jr3y717}{Ohio University}, Athens, Ohio~45701, USA}

\author{J.~Kim~\orcid{0009-0003-9488-0247}}
\email{jane.kim@anl.gov}
\affiliation{Physics Division, \href{https://ror.org/05gvnxz63}{Argonne National Laboratory}, Lemont, Illinois~60439, USA}
\affiliation{Department of Physics and Astronomy, \href{https://ror.org/01jr3y717}{Ohio University}, Athens, Ohio~45701, USA}

\author{T.~Zhao~\orcid{0000-0003-4704-0109}}
\email{tianqi.zhao@berkeley.edu}
\affiliation{Institute for Nuclear Theory, \href{https://ror.org/00cvxb145}{University of Washington}, Seattle, Washington~98195, USA}
\affiliation{Network for Neutrinos, Nuclear Astrophysics, and Symmetries (N3AS), \href{https://ror.org/01an7q238}{University of California, Berkeley}, California~94720, USA}

\author{C.~Drischler~\orcid{0000-0003-1534-6285}}
\email{drischler@ohio.edu}
\affiliation{Department of Physics and Astronomy, \href{https://ror.org/01jr3y717}{Ohio University}, Athens, Ohio~45701, USA}
\affiliation{\href{https://ror.org/03r4g9w46}{Facility for Rare Isotope Beams}, \href{https://ror.org/05hs6h993}{Michigan State University}, East Lansing, Michigan~48824, USA}

\date{\today}
\preprint{N3AS-26-017, INT-PUB-26-029}

\begin{abstract} 

We present constraints on the nuclear equation of state (EOS) from microscopic asymmetric matter calculations at zero temperature based on chiral nucleon-nucleon and three-nucleon interactions. The constraints include the saturation point, the isospin dependence of the incompressibility, and the symmetry energy, as well as the crust-core transition density of neutron-star matter. To quantify and propagate correlated uncertainties from noisy many-body calculations to derived observables, we introduce \texttt{GPDiff}, an efficient \texttt{JAX}-based Python package for multivariate Gaussian process (GP) regression with automatic differentiation. After training, \texttt{GPDiff} enables joint predictions of the EOS and derivatives of arbitrary order with respect to the input variables, including mixed partial derivatives. In this initial application, we analyze recent high-order many-body perturbation theory calculations of asymmetric matter up to about twice saturation density and explore nonstationary change-surface kernels, a class of input-dependent kernels, for modeling the EOS. \texttt{GPDiff} is broadly applicable to microscopic nuclear EOS calculations at zero and finite temperature and provides a versatile package for GP-based uncertainty quantification and inference of the nuclear EOS.

\end{abstract}

\maketitle

\section{Introduction}
\label{sec:intro}

Infinite nuclear matter, an idealized system of interacting nucleons without Coulomb and surface effects, has been an important laboratory for benchmarking microscopic nuclear forces in medium~\cite{Ekstrom:2015rta,Ekstrom:2017koy,Drischler:2017wtt,Jiang:2022oba,Jiang:2022tzf,Drischler:2024ebw}. 
Its equation of state (EOS) provides essential constraints on the properties of strongly interacting matter and is thus essential for understanding atomic nuclei, heavy-ion collisions, and compact objects such as neutron stars~\cite{Drischler:2021kxf,Sorensen:2023zkk,MUSES:2023hyz,Chatziioannou:2024jsr,Agarwal:2025ezo}. 

Substantial advances in chiral effective field theory (EFT) for nuclear forces~\cite{Epelbaum:2008ga,Machleidt:2011zz,Hammer:2019poc, Epelbaum:2019kcf} and in many-body frameworks, including high-order many-body perturbation theory (MBPT)~\cite{Drischler:2017wtt,Drischler:2026vdm}, quantum Monte Carlo (QMC) methods~\cite{Lonardoni:2019ypg,Arthuis:2022ixv,Fore:2024exa,Lovato:2026erx}, and emulators for uncertainty quantification~(UQ)~\cite{Duguet:2023wuh,Melendez:2022kid,Drischler:2022ipa,Cook:2024toj,Frame:2017fah}, have provided increasingly stringent constraints on low-density nuclear matter. 
In particular, recent calculations of symmetric nuclear matter (SNM) and pure neutron matter (PNM) have provided insights into nuclear saturation properties and, when combined with empirical expansions in the isospin asymmetry, constrain the EOS of neutron-rich matter with important implications for the structure of neutron stars~\cite{Hebeler:2013nza,Lynn:2015jua,Holt:2016pjb,Drischler:2020fvz,Huth:2020ozf,Drischler:2021bup,Huth:2021bsp,Tews:2024owl,Alp:2025wjn,Drischler:2026vdm}.
In parallel, explicit calculations of isospin asymmetric matter allow for direct tests of these expansions while further constraining the infinite matter EOS as a function of the density and isospin asymmetry, both at zero and finite temperature~\cite{Drischler:2013iza,Drischler:2015eba,Wellenhofer:2016lnl,Somasundaram:2020chb,Wen:2020nqs,Keller:2020qhx,Keller:2022crb}.

Gaussian processes~(GPs)~\cite{Mackay:2003information,rasmussen2006gaussian,Murphy:2012abc,gelman2013bayesian} are a well-established nonparametric method for regression and classification in machine learning\footnote{%
For connections between GPs and infinite-width neural networks, see Refs.~\cite{Neal1996,Williams1996ComputingWI,lee2018deep,matthews2018gaussian}; for related work in nuclear physics, see Ref.~\cite{Sundberg:2025lbu}.%
} 
and have become popular statistical tools for EOS modeling across a wide range of densities with quantified uncertainties.
For example, GPs have been used to estimate truncation errors from finite-order EFT or many-body expansions~\cite{Drischler:2020hwi,Drischler:2020yad,Svensson:2025jde,Gottling:2025ohe}; 
interpolate many-body calculations to constrain derived quantities such as the speed of sound~\cite{Drischler:2020hwi,Drischler:2020yad,Keller:2022crb}; 
facilitate Bayesian model mixing to combine EOS constraints from different density regions into globally predictive composite models~\cite{Semposki:2024vnp,Semposki:2025etb}; 
and provide model-agnostic priors for EOS inference from multimessenger observations of neutron stars (e.g., Refs.~\cite{Gorda:2026rzm,Legred:2026zok,Finch:2025bao,Gorda:2025aiu,Legred:2025aar,Ng:2025wdj,Golomb:2024mmt,Essick:2023fso,Legred:2023als,Legred:2021hdx,Landry:2018prl}).

Going forward, it will be important to explore more systematically how inferred EOS constraints depend on modeling choices, such as the GP kernel, which encodes prior information about the EOS function space, and how these choices can be informed by physics. 
While advanced kernel design is well established in machine learning~\cite{duvenaud_PhD_2014}, e.g., to incorporate information on data trends and patterns, it has only recently begun to see broader use in nuclear physics.
Recent applications to neutron-star matter include modified GPs designed to model nontrivial features in the speed of sound~\cite{Mroczek:2023zxo} and nonstationary changepoint kernels used in Bayesian model mixing of chiral EFT and perturbative quantum chromodynamics (pQCD) calculations~\cite{Semposki:2024vnp,Semposki:2025etb}.

Motivated by these considerations, we have developed a general-purpose Python package for multivariate GP regression, named \gpdiff, which enables theoretical uncertainties in the EOS to be quantified and propagated to derived quantities for arbitrary differentiable kernels.\footnote{%
Appendix~\ref{app:other_gp_libs} contains a listing of other GP-based software packages, including packages using Google's \texttt{JAX}.%
}
For example, once trained on a set of many-body calculations for the energy per particle based on various chiral interactions, \gpdiff can efficiently evaluate derivatives to constrain derived observables, such as the pressure, with quantified correlated uncertainties.
Built on Google's \texttt{JAX}~\cite{jax2018github}, \gpdiff leverages automatic differentiation and just-in-time compilation, among other high-performance computing optimizations. 
While addressing use cases similar to those of the earlier Python package \texttt{gptools}~\cite{Chilenski_2015_gptools}, which, to our knowledge, is no longer actively maintained, \gpdiff provides an independent, more flexible \texttt{JAX}-based implementation tailored to EOS UQ and inference.

In this work, we demonstrate \gpdiff's efficacy for analyzing microscopic calculations of asymmetric nuclear matter using a higher-dimensional generalization of the nonstationary change-point kernels recently studied for Bayesian model mixing in Ref.~\cite{Semposki:2025etb}. 
We present results for several low-density EOS parameters, including the incompressibility of SNM and the symmetry energy. 
In particular, we constrain the EOS parameter $K_\tau$, which governs the isospin dependence of the incompressibility $K$ at leading order~\cite{Piekarewicz:2008nh,Piekarewicz:2009gb}. 
Improving experimentally informed constraints on $K_\tau$ is currently the goal of the scheduled FRIB experiment ``The Isoscalar Giant Monopole Resonance in \isotope[132]{Sn}: Implications on the Nuclear Incompressibility'' (PAC \#21056). 
We also constrain properties of low-density neutron-star matter, such as the EOS and the crust-core transition density, below which matter is nonuniform. 
While our constraints are based on recent high-order MBPT calculations of asymmetric matter~\cite{Drischler:2026vdm}, \gpdiff is broadly applicable to many-body calculations of the nuclear EOS at zero and finite temperature.

The remainder of this paper is organized as follows.
Section~\ref{sec:framework} describes our statistical framework, including the developed Python library \gpdiff and the proposed deviation model for noisy microscopic EOS calculations.
Together with technical documentation and the EOS results presented here, \gpdiff will be made publicly available on GitHub to allow the practitioner to apply and extend it in their research.
In Sec.~\ref{sec:results}, we apply \gpdiff to recent asymmetric matter calculations and present our main results for constraining the symmetry energy and other EOS parameters, as well as neutron-star matter properties, such as the EOS, composition, and crust-core transition density.
Section~\ref{sec:summary_outlook} concludes with a summary and outlook.
Appendix~\ref{app:other_gp_libs} gives an overview of existing libraries for GP regression and classification; 
Appendix~\ref{app:UP_nsEOS} provides further details on the uncertainty propagation to neutron-star matter;
and Appendix~\ref{app:add_results} contains additional results based on a different kernel choice that supplement Sec.~\ref {sec:results}.
We use natural units in which $\hbar = c = 1$ and typeset vectors and matrices in boldface.

\section{GP-based framework}
\label{sec:framework}

In this section, we describe our GP-based framework for modeling cold, low-density matter based on microscopic EOS calculations. 

\subsection{GPs and their implementation in \gpdiff}

To keep the article self-contained, we summarize the key features of GPs relevant to our EOS analysis, including kernels, mean functions, and differentiation, and discuss their implementation in \gpdiff.
For a comprehensive introduction to GPs and their applications, we refer the reader to Refs.~\cite{Mackay:2003information,rasmussen2006gaussian,Murphy:2012abc,gelman2013bayesian}; 
for an introduction in the context of correlated EFT truncation errors specifically, see Section~II.B of Ref.~\cite{Melendez:2019izc}.

A GP is a collection of random variables, any finite number of which have joint Gaussian distributions~\cite{rasmussen2006gaussian}.
A one-dimensional GP $f(\vb{x})$, with multiple inputs collected in the $d$-dimensional vector $\vb{x}$, is denoted by
\begin{equation}
f\left( \vb{x} \right) \sim \mathcal{GP}\left( \mu (\vb{x}), k \left( \vb{x}, \vb{x}' \right) \right) \,,
\end{equation}
and is fully determined by its mean function
\begin{equation}
\mu\left( \vb{x} \right) = \mathbb{E} \left[ f\left( \vb{x} \right) \right] \,,
\end{equation}
and kernel, also called the covariance function,
\begin{equation}
k\left( \vb{x}, \vb{x}' \right) = \mathbb{E}\left[ \left( f\left( \vb{x} \right) - \mu\left( \vb{x} \right) \right) \left( f\left( \vb{x}' \right) - \mu\left( \vb{x}' \right) \right)  \right] \,.
\end{equation}
In our zero-temperature EOS applications, we model the energy per particle as a GP, $f(\vb{x}) = E(\vb{x})/A$, with $d=2$ inputs, where $\vb{x} = (\diso, n)$ denotes the baryon density $n$ and isospin asymmetry $\diso$, both continuous variables.
In practice, however, the EOS is only evaluated at a finite set of $N$ observation (or training) points, which we collect as $\vb{x}_t = \{\vb{x}_{i}\}_{i=1}^N$. For example, evaluating PNM ($\diso =1$) and SNM ($\diso=0$) at the saturation density $n_0$ corresponds to $N=2$.
The associated function values, $\vb{f}_t = \{f(\vb{x}_{i})\}_{i=1}^N$, are modeled as random variables following a multivariate normal distribution,\footnote{%
We follow the standard statistical notation, denoting the covariance matrix by $\vb{K}$, which should not be confused with the incompressibility of SNM, $K$.
} 
\begin{equation} \label{eq:gp_norm_dist}
    \vb{f}_t \given \vb{x}_t \sim \normal(\vb*{\mu}_t, \vb{K}_t)\,,
\end{equation}
where the mean vector $\vb*{\mu}_t$ has the components $\mu_{i} = \mu(\vb{x}_{i})$ and the covariance (or kernel) matrix $\vb{K}_t$ has elements $(\vb{K}_t)_{ij} = k(\vb{x}_{i}, \vb{x}_{j})$.

By modeling the EOS as a GP with a specified mean function and kernel, we impose a \emph{prior} probability distribution over functions that, before calibration to data, can represent a wide range of behaviors consistent with our prior knowledge of the EOS's smoothness and global trends, rather than restricting it to a single parametric form.
The chosen mean function encodes our prior expectation of the systematic trends in the data (i.e., microscopic EOS calculations), while the covariance structure of the deviations from this expectation is specified by the kernel.
Calibration to data then yields a posterior distribution over functions, as described in the next subsections.

\subsubsection{Kernels}
 
A valid kernel must be symmetric and positive semi-definite, which ensures that the associated matrix $\vb{K}$ defines a proper covariance matrix.
The parameters of a kernel, such as length scales, are called hyperparameters.
We collect them in $\boldsymbol{\theta}$ and calibrate their values to data.

There are two broad classes of kernels~\cite{Genton:2002,duvenaud_PhD_2014,Murphy:2012abc}.
Stationary kernels are translation invariant, i.e., $k(\vb{x}, \vb{x}') = k(\vb{x} - \vb{x}')$, so the covariance between function values depends on their relative separation and not on their absolute location.\footnote{%
If a stationary kernel is also isotropic, then it depends only on the distance between inputs, $k(\vb{x}, \vb{x}') = k(\norm{\vb{x} - \vb{x}'})$.%
}
Examples include the constant, radial basis function (RBF), Mat\'ern, and rational quadratic (RQ) kernels. 
In contrast, nonstationary kernels allow the covariance to vary with the absolute position in input space.
Examples include the polynomial and change-point (CP) kernels. 

The most widely used kernel is the stationary RBF kernel, also known as the squared-exponential (SE) kernel:
\begin{equation} \label{eq:rbf_kernel}
k_\text{RBF} \left( \vb{x}, \vb{x}' ; \vb*{\theta}\right)  = \sigma_f^2 \exp \left[-\frac{1}{2}
 \left( \vb{x} - \vb{x}' \right)^{\intercal} \vb*{\Lambda}_{\vb*{\theta}}^{-1} \left( \vb{x} - \vb{x}' \right)
 \right] \,.
\end{equation}
It is parameterized by the signal variance $\sigma_f^2 > 0$ and a set of length scales for each input dimension, collected in the diagonal matrix $\vb*{\Lambda}_{\vb*{\theta}} = \diag(\lambda_1^2, \lambda_2^2, \ldots, \lambda_d^2)$.
Despite its popularity, the RBF kernel can be too restrictive for EOS modeling across a wide range of densities because it is infinitely differentiable, implying global smooth, stationary behavior with a single length scale in each dimension.

\begin{figure*}[tb]
    \includegraphics[width=\textwidth]{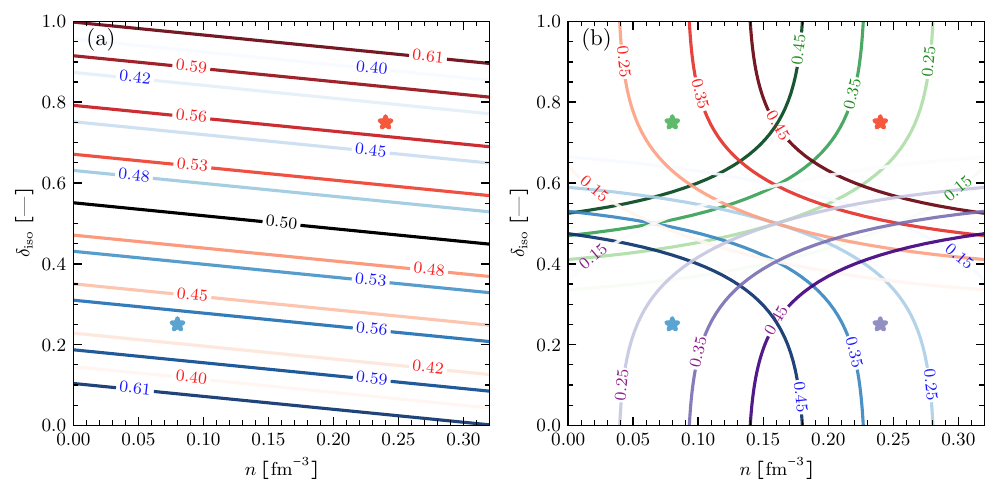} 
    \caption{%
    Illustrations of two-dimensional change-surface kernels (e.g., the $n$--$\diso$ plane) with two centers (panel~a) and four centers (panel~b), as indicated by the colored stars.
    For illustration, we assume that the weight functions in Eq.~\eqref{eq:multi-cp_kernel} have the form of normal distributions, with their means located at the colored stars and identity covariance matrices.
    The solid lines depict the contours (i.e., level sets) for which the weights of the corresponding kernels are as annotated.
    For example, in panel~(a), the so-called change point would correspond to the point where the contour line corresponding to $w = 0.5$ and the line connecting the two centers intersect.
    In panel~(b), consider the green center and the associated region encompassed by, e.g., the $w = 0.45$ contour line, which corresponds to low-density neutron-rich matter.
    This center separates characteristic EOS features in that region from those of neutron-rich matter at higher densities ($n \gtrsim n_0$) and from nearly SNM, where nuclear saturation emerges.%
    }
    \label{fig:illustration_multi-cp_kernel}
\end{figure*}

Individual kernels can also be combined through addition, multiplication, or scaling to capture more complex structure in the output space. 
In this regard, nonstationary change-surface (CS) kernels, which combine two or more kernels via input-dependent weight functions, allow the covariance structure to vary across regions of the input space and are therefore of particular interest for EOS inference~\cite{Semposki:2025etb}. 
Here, we consider the subclass of CS kernels~\cite{Herlands:2015,Wilson:2011} with smooth Gaussian weight (or gating) functions,
\begin{subequations}\label{eq:multi-cp_kernel}
\begin{align}
    k_{\mathrm{CS}}(\vb{x},\vb{x}')
    &= \sum_i \sigma_i(\vb{x})\, k_i(\vb{x},\vb{x}')\, \sigma_i(\vb{x}') \,, \\
    \sigma_i(\vb{x})
    &= \frac{a_i}{Z}\exp\!\left[-(\vb{x}-\vb{c}_i)^{\intercal}\vb{B}^{-1}(\vb{x}-\vb{c}_i)\right] \,,
\end{align}
\end{subequations}
where $\vb{c}_i$ denotes the center of the $i$th weight function $\sigma_i(\vb{x})$, and $Z$ is determined such that $\sum_i \sigma_i(\vb{x})=1$ for all $\vb{x}$. 
For illustration, throughout this article we set all $a_i$ to be equal (e.g., $a_i=1)$, take a common diagonal smearing matrix $\vb{B}=\diag(b_1^2,b_2^2,\ldots,b_d^2)$, and adjust only the widths $b_i>0$ and the centers $\vb{c}_i$.

It is worth noting that, when combining only two kernels, Eq.~\eqref{eq:multi-cp_kernel} reduces to the multivariate version of the CP kernel studied in Ref.~\cite[see Eq.~(29)]{Semposki:2025etb} for Bayesian mixing of EOS models:
\begin{subequations} \label{eq:mcp_kernel}
\begin{align}
k_{\mathrm{CP}} \left( \vb{x}, \vb{x}' \right) &= \sigma \left( \vb{x} \right)  k_1 \left(\vb{x}, \vb{x}' \right) \sigma \left( \vb{x}' \right)  \notag\\
& \quad + \left( 1 - \sigma \left( \vb{x} \right) \right) k_2 \left( \vb{x}, \vb{x}' \right)  \left( 1 - \sigma \left( \vb{x}' \right) \right) \,, \\
 \sigma (\vb{x}) &= \left[1 + \exp \left[-\vb{w} \cdot \left(\vb{x}-\vb{x}_c\right) \right] \right]^{-1} \,, \label{eq:mcp_weight}
\end{align}
\end{subequations}
where the mid-point $\vb{x}_c$ and normal vector $\vb{w}$,
\begin{align}
    \vb{x}_c = \frac{\vb{c}_1+\vb{c}_2}{2} \,, & & 
    \vb{w} = 2 \vb{B}^{-1} \left(\vb{c}_1-\vb{c}_2\right) \,, \label{eq:params_sigmoid}
\end{align}
define the sigmoidal weight function $\sigma(\vb{x})$.\footnote{%
Note that $a$ does not contribute to the weight function~\eqref{eq:mcp_weight}, because setting $a_1 = a_2$ renders $a$ a global prefactor that is absorbed into the overall normalization.
} 
Its level sets (contours) are hyperplanes defined by $\vb{w}\cdot \vb{x} = \text{constant}$, which are straight lines in two dimensions. The level set $\sigma(\vb{x}) = 0.5$ defines a hyperplane passing through $\vb{x}_c$, referred to as the change surface (or change point in one dimension), given by
\begin{equation} \label{eq:def_change_boundary}
    \vb{w} \cdot \vb{x} = \vb{w} \cdot \vb{x}_c \equiv \vb{c}_1^{\intercal} \vb{B}^{-1}\vb{c}_1 - \vb{c}_2^{\intercal} \vb{B}^{-1}\vb{c}_2 \,.
\end{equation}
In the limit $\vb{c}_2 \to \vb{c}_1$, Eq.~\eqref{eq:def_change_boundary} is trivially satisfied; i.e., both constituent kernels contribute equally for all $\vb{x}$, and thus $\sigma(\vb{x}) \equiv \tfrac{1}{2}$.

For illustration, Fig.~\ref{fig:illustration_multi-cp_kernel} shows the CS kernel~\eqref{eq:multi-cp_kernel} for two (panel~a) and four (panel~b) centers in the $n$--$\diso$ plane ($d=2$). 
The stars indicate the locations of the centers for the two illustrative choices of $\vb{B}$ shown in panels~(a) and~(b), respectively.
As discussed, the change surface in panel~(a) is a straight line (i.e., the black line). 
The remaining solid lines show contour levels at the annotated values and are associated with centers of the corresponding color. 
Figure~\ref{fig:illustration_multi-cp_kernel} illustrates that CS kernels of the form of Eq.~\eqref{eq:mcp_kernel} enable the construction of composite kernels designed to represent different features of the EOS in its input space, e.g., $(\diso,n)$.
For example, consider the green center and its associated contour lines in Fig.~\ref{fig:illustration_multi-cp_kernel}(b). 
This center separates characteristic EOS features of neutron-rich matter at low densities ($n \lesssim n_0$) from those of neutron-rich matter at higher densities ($n \gtrsim n_0$) and from nearly SNM, where nuclear saturation emerges.
In our implementation, the two degrees of freedom are the centers $\vb{c}_i$, which determine the locations of these features, and the global smearing parameters $b$ in each dimension, which control the smoothness of the transition between the kernels.
For $b \to 0$, these features are assumed isolated since $\sigma(\vb{x})$ approaches the Heaviside step function centered at $\vb{x}_c$, resulting in a discontinuous transition. 
For $b \to \infty$, both kernels contribute equally across the entire input space. 
For intermediate values of $b$, the transition between the kernels is more localized but still smooth.

\subsubsection{Predictions and calibration}

In practical applications, such as many-body EOS calculations, observations are subject to uncertainties. 
That means, instead of observing the latent function directly, one observes noisy realizations of it. 
Nevertheless, we can infer the latent function and predict its values at new input points, as described below.

Let $\vb{x}_t = \{\vb{x}_1, \dots, \vb{x}_N\}$ be the set of (observed) training inputs, with corresponding observations $\vb{y}_t$. 
These observations are modeled here as $\vb{y}_t = \vb{f}_t + \vb{\epsilon}$ with $\vb{\epsilon} \sim \mathcal{N}(\vb{0}, \vb{C}_t)$,
where $\vb{f}_t \equiv \vb{f}(\vb{x}_t)$ are the latent function values at $\vb{x}_t$, and $\vb{C}_t$ is the covariance matrix describing the observational uncertainties.
Sometimes only the diagonal components of $\vb{C}_t$ are considered, e.g., in the Python package \texttt{scikit-learn}~\cite{scikit-learn}, but here we retain the full covariance matrix to account for correlated uncertainties in the input data.
Our goal is then to predict the latent function at a set of $N_*$ evaluation points $\vb{x}_* = \{\vb{x}_{*1}, \vb{x}_{*2}, \dots, \vb{x}_{*N_*}\}$, with corresponding values $\vb{f}_* \equiv \vb{f}(\vb{x}_*)$. 
The joint distribution of the observations and the latent function at the evaluation points is given by:
\begin{equation}
\begin{bmatrix}
\vb{y}_t \\
\vb{f}_*
\end{bmatrix}
\sim
\mathcal{N}\left(
\begin{bmatrix}
\vb*{\mu}_{t} \\
\vb*{\mu}_{*}
\end{bmatrix},
\begin{bmatrix}
\vb{K}_{tt} + \vb{C}_t & \vb{K}_{t*} \\
\vb{K}_{*t} & \vb{K}_{**}
\end{bmatrix}
\right),
\end{equation}
where $\vb*{\mu}_t \equiv \vb*{\mu}(\vb{x}_t)$ and $\vb*{\mu}_* \equiv \vb*{\mu}(\vb{x}_*)$ are the mean functions evaluated at the training and test inputs, respectively, and the covariance matrices have elements $[\vb{K}_{tt}]_{ij} = k(\vb{x}_{t,i}, \vb{x}_{t,j}; \vb*{\theta})$, $[\vb{K}_{t*}]_{ij} = k(\vb{x}_{t,i}, \vb{x}_{*j}; \vb*{\theta})$, and $[\vb{K}_{**}]_{ij} = k(\vb{x}_{*i}, \vb{x}_{*j}; \vb*{\theta})$, generated by the kernel $k$ with kernel hyperparameter $\vb*{\theta}$.
Hence, the kernel can be seen as a prior on the covariance structure, which is updated by conditioning on the observed data $\vb{y}_t$, with their uncertainties encoded in $\vb{C}_t$.

Conditioning on the training data $(\vb{x}_t,\vb{y}_t)$, one can show that the posterior distribution for the latent function values at the test points is the multivariate normal
\begin{equation} \label{eq:gp_posterior_latent}
\vb{f}_* \given \vb{x}_*, \vb{x}_t, \vb{y}_t, \vb*{\theta}
\sim \mathcal{N}\left( \vb{m}_{*|t}, \vb{K}_{**|t} \right) \,,
\end{equation}
with the predictive mean and covariance:
\begin{align}
\vb{m}_{*|t}
&= \vb*{\mu}_{*} + \vb{K}_{*t} \left[ \vb{K}_{tt} + \vb{C}_t \right]^{-1} (\vb{y}_t - \vb*{\mu}_{t}) \,, \label{eq:gp_pred_mean} \\
\vb{K}_{**|t}
&= \vb{K}_{**}  - \vb{K}_{*t} \left[ \vb{K}_{tt} + \vb{C}_t \right]^{-1} \vb{K}_{t*} \label{eq:gp_pred_cov}\,.
\end{align}
Comparing Eq.~\eqref{eq:gp_norm_dist} with Eqs.~\eqref{eq:gp_pred_mean} and~\eqref{eq:gp_pred_cov} reveals that the data-agnostic GP prior~\eqref{eq:gp_norm_dist} is updated by the data-informed second terms in the predictive mean~\eqref{eq:gp_pred_mean} and covariance~\eqref{eq:gp_pred_cov}. 
For example, for the RBF kernel, evaluation points far from the data support (in terms of the kernel's correlation length) cause $\vb{K}_{*t}$ to approach the zero matrix and thus the GP reverts to its prior mean and covariance structure in Eq.~\eqref{eq:gp_norm_dist}.

In general, the hyperparameters $\vb*{\theta}$ are calibrated to the observed data.
Given observations $\vb{y}_t$ at inputs $\vb{x}_t$, the posterior for the hyperparameters is
\begin{equation} \label{eq:hyperposterior}
  \pr(\vb*{\theta} \given \vb{x}_t, \vb{y}_t) \propto \pr(\vb{y}_t \given \vb{x}_t, \vb*{\theta})\,\pr(\vb*{\theta}) \,,
\end{equation}
with the so-called log marginal likelihood,
\begin{multline} \label{eq:marg_like}
\ln \pr(\vb{y}_t \given \vb{x}_t, \vb*{\theta})
=
-\frac12 \ln|\vb{K}_{tt}+\vb{C}_t|
\\
-\frac{1}{2}\tilde{\vb{y}}_t^{\intercal} \left (\vb{K}_{tt}+\vb{C}_t \right)^{-1} \tilde{\vb{y}}_t
 \,,
\end{multline}
which is obtained by marginalizing out the typically unobserved latent function, $\vb{f}_t$.
In Eq.~\eqref{eq:marg_like}, we have used the short-hand notation $\tilde{\vb{y}}_t = \vb{y}_t-\vb*{\mu}_t$ and dropped the additive constant independent of $\vb*{\theta}$.
The hyperposterior~\eqref{eq:hyperposterior} can be sampled using Monte Carlo (MC) methods, which allows one to explore a wide function space informed by the observed data, and predictions can then be made by marginalizing out the hyperparameters:
\begin{equation} \label{eq:marg_pred}
  \pr(\vb{f}_* \given \vb{x}_t, \vb{y}_t)
  = \int \pr(\vb{f}_* \given \vb{x}_t, \vb{y}_t, \vb*{\theta})
  \pr(\vb*{\theta} \given \vb{x}_t, \vb{y}_t) \dd{\vb*{\theta}}.
\end{equation}
This process, however, is computationally demanding.
Hence, we follow here instead the usual maximum a~posteriori (MAP) approach, in which only the maximum of Eq.~\eqref{eq:hyperposterior}, $\vb*{\theta}_{\mathrm{MAP}}$, is used as a point estimate, corresponding to $\delta$-distribution $\pr(\vb*{\theta} \given \vb{x}_t, \vb{y}_t) = \delta(\vb*{\theta}-\vb*{\theta}_{\mathrm{MAP}})$ in Eq.~\eqref{eq:marg_pred}.
In \gpdiff, this optimization is performed using stochastic gradient descent via the \texttt{optax} library~\cite{deepmind2020jax}.

\subsubsection{GP differentiation}
\label{sec:gp_differentiation}

One of the key features that makes GPs attractive for EOS inference (and thermodynamics in general) is that they are closed under linear operations, such as differentiation, integration, and linear combinations.
That means any linear operation applied to a GP results in another GP.
In this work, we are mainly interested in derivatives of the GPs trained on many-body calculations of the energy per particle to obtain derived quantities such as the pressure.

For brevity, we assume in the following that a scalar-valued function of a one-dimensional input $x$ is modeled as a GP.
The generalization to vector inputs $\vb{x}$ can be found, e.g., in Ref.~\cite{lee2026thesis}.
Since derivatives of GPs remain GPs, if they exist (see below), $f(x)$ and its derivatives $f'(x), f''(x), \dots, f^{(n)}(x)$ are jointly normally distributed.
This property enables simultaneous prediction of the EOS and its relevant derivatives while consistently quantifying and propagating uncertainties in the input data.
For illustration, let us only consider the joint distribution of the noisy observations and the $n$-th derivative at $\vb{x}_*$, which is given by:
\begin{align}
\begin{bmatrix} 
\vb{y}_t \\ \vb{f}_*^{(n)} 
\end{bmatrix} 
\sim \mathcal{N} \left(
\begin{bmatrix}
\vb*{\mu}_t \\
\vb*{\mu}_*^{(n)}
\end{bmatrix},
\begin{bmatrix}
\vb{K}_{tt} + \vb{C}_t & \vb{K}_{t*}^{(0,n)} \\
\vb{K}_{*t}^{(n,0)} & \vb{K}_{**}^{(n,n)} \\
\end{bmatrix}
\right) \,,
\end{align}
where $\vb*{\mu}_*^{(n)} \equiv \vb*{\mu}^{(n)}(\vb{x}_*)$ is the $n$-th derivative of the mean function evaluated at the test inputs, and the covariance matrices have elements $[\vb{K}_{t*}^{(0,n)}]_{ij} = k^{(0,n)}(\vb{x}_{t,i}, \vb{x}_{*j}; \vb*{\theta})$, $[\vb{K}_{*t}^{(n,0)}]_{ij} = k^{(n,0)}(\vb{x}_{*i}, \vb{x}_{t,j}; \vb*{\theta})$, and $[\vb{K}_{**}^{(n,n)}]_{ij} = k^{(n,n)}(\vb{x}_{*i}, \vb{x}_{*j}; \vb*{\theta})$, generated by the differentiated kernel
\begin{equation} \label{eq:kernel_deriv}
    k^{(a,b)} \left( x, x' \right) = \frac{\partial^{a + b}}{\partial {x}^a \partial {(x')}^b} k \left( x, x' \right) \,.
\end{equation}
Similar to the derivation of Eq.~\eqref{eq:gp_posterior_latent}, it follows that, given the training data, the distribution of $\vb{f}^{(n)}_*$ at test locations $\vb{x}_*$ is:
\begin{equation}  \label{eq:gp_deriv_posterior_latent}
\vb{f}^{(n)}_* \given \vb{x}_*, \vb{x}_t, \vb{y}_t  \sim \mathcal{N} \left( \vb{m}_{*|t}^{(n)}, \vb{K}_{**|t}^{(n,n)} \right)\,,
\end{equation}
with the predictive mean and covariance,
\begin{align}
\vb{m}_{*|t}^{(n)} &= \vb*{\mu}^{(n)}_{*} + \vb{K}^{(n,0)}_{*t} \left[ \vb{K}_{tt} + \vb{C}_t \right]^{-1} \left( \vb{y}_t - \vb*{\mu}_t \right) \,, \label{eq:mean_deriv_gp} \\
\vb{K}_{**|t}^{(n,n)} &= \vb{K}^{(n,n)}_{**} - \vb{K}^{(n,0)}_{*t} \left[ \vb{K}_{tt} + \vb{C}_t \right]^{-1} \vb{K}^{(0,n)}_{t*} \,.\label{eq:cov_deriv_gp}
\end{align}

A few comments are in order. 
Note that in Eq.~\eqref{eq:mean_deriv_gp}, the mean of the $n$-th derivative GP is given by the $n$-th derivative of the mean function plus corrections from the covariance structure.
Similarly, the covariance is given by the differentiated kernel matrix plus corrections in Eq.~\eqref{eq:cov_deriv_gp}.
This property ensures that the global trend encoded in $\mu(x)$, together with its covariance structure, is consistently propagated to all derivatives. 
Hence, the existence of predictions for $n$-th order derivatives of the original GP depends on the differentiability of both the mean function and the kernel.
Specifically, for the $n$-th derivative to exist, the $\mu(x)$ must be at least $n$-times differentiable, and $k(x,x')$ must be at least $n$-times differentiable with respect to both arguments over the entire input domain of interest.
For example, for the second-derivative GP to exist, the kernel must admit all partial derivatives with $a,b \leqslant n = 2$, as can be seen in Eq.~\eqref{eq:kernel_deriv}.
These conditions are not satisfied by all choices of mean functions and kernels. 
For example, the Mat{\'e}rn kernel (with finite degree of freedom) is not smooth, which restricts the order of derivatives that can be computed.

While the mean function and kernel derivatives can be derived and implemented manually, e.g., as in \texttt{gptools}, \gpdiff uses \texttt{SymPy} to evaluate the derivatives of the mean function symbolically,\footnote{%
Automatic differentiation could also be used. 
The symbolic \texttt{SymPy} implementation may be advantageous, e.g., when mean functions are expressed in terms of physics-informed basis expansions.
} 
and uses automatic differentiation built into Google's \texttt{JAX} to implement the derivative of the kernel.
This feature enables arbitrary (partial) derivatives of the EOS with respect to any input variable to be computed straightforwardly at machine precision for a wide range of mean functions and kernels.
Moreover, \gpdiff enables joint random sampling of the original GP and its derivatives, with correlated uncertainties in the input data propagated.
We see therein a significant advantage over other methods, including finite differencing, especially for noisy data sets.

\subsection{Microscopic training data}
\label{sec:micro_data}

\begin{figure*}[p]
        \includegraphics[width=1\linewidth]{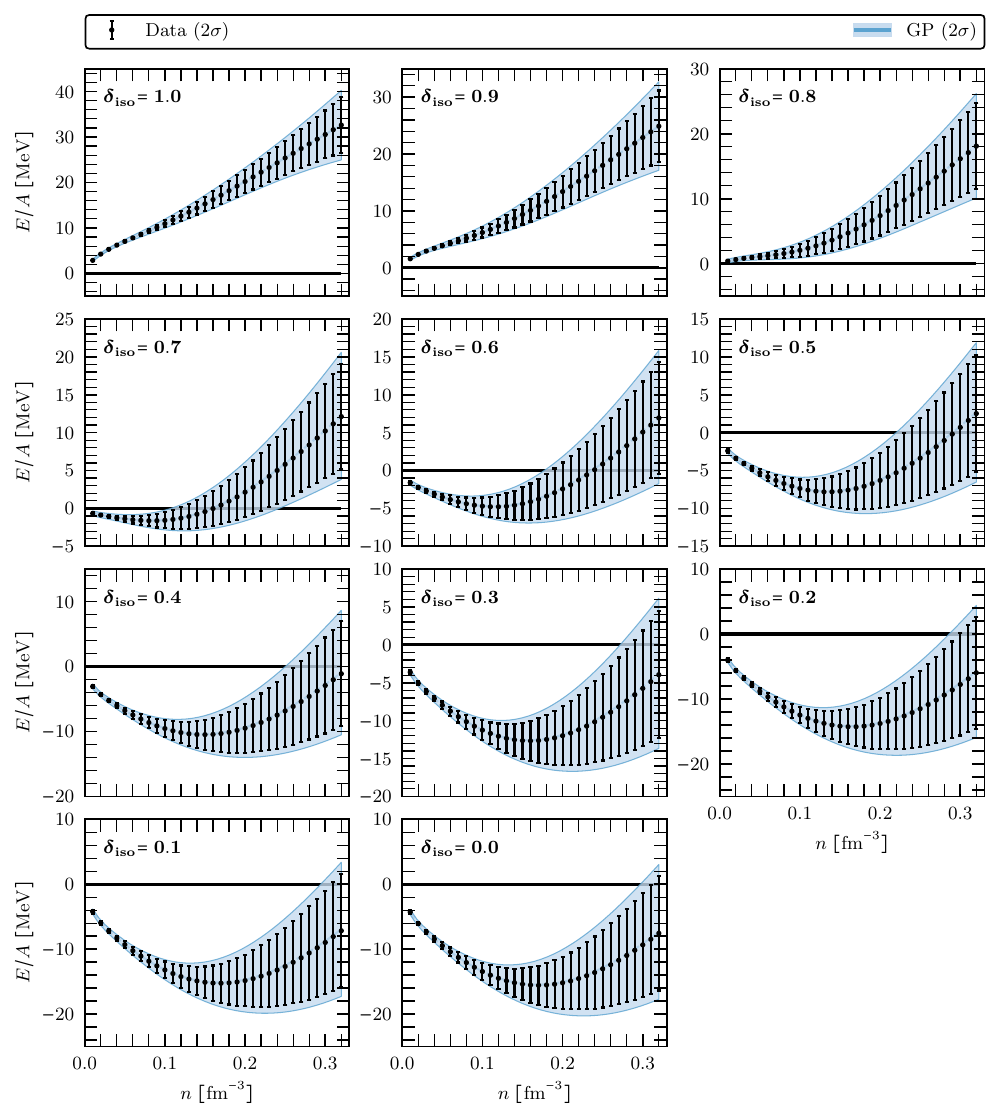}
        \caption{%
    Explicit asymmetric matter calculations used for training the GPs.
    In Ref.~\cite{Drischler:2026vdm}, the energy per particle as a function of the density $n$, sampled at 11 equidistant isospin asymmetries $\diso$ between PNM ($\diso=1$) and SNM ($\diso=0$), was calculated for the six Hebeler et al.\ interactions using MBPT up to fourth order at the normal-ordered two-body level.
    The black dots depict the sample mean values and the error bars depict twice the standard deviations across these interactions.
    The blue uncertainty bands correspond to our GP-based EOS model at the $2\sigma$ confidence level.
    See the main text for details.%
 }
    \label{fig:anm_mbpt}
\end{figure*}

We examine microscopic calculations of the zero-temperature energy per particle $E(\diso, n)/A$ as a function of the nucleon density $n = \iso{n}{n} + \iso{n}{p}$ and isospin asymmetry $\diso = (\iso{n}{n}-\iso{n}{p})/(\iso{n}{n}+\iso{n}{p})$, where $\iso{n}{p}$ and $\iso{n}{n}$ are the proton and neutron densities, respectively.\footnote{%
	The proton fraction and asymmetry are related via $\diso = 1 - 2x$. Hence, we use both variables interchangeably.%
}
We calibrate a GP to these calculations using \gpdiff, while all other observables, such as the pressure, are derived from this GP.
Section~\ref{sec:discrepancy_model} discusses the details of this calibration.

Specifically, we consider the explicit MBPT calculations of asymmetric matter based on chiral nucleon-nucleon (NN) and three-nucleon (3N) interactions in Ref.~\cite{Drischler:2026vdm} as the training set for EOS analysis, although this is merely a choice as our framework is more broadly applicable to microscopic many-body calculations.
Reference~\cite{Drischler:2026vdm} computed $E(\diso,n)/A$ for $n \leqslant 0.32 \fmiq$ ranging from PNM ($\diso=1$) to SNM ($\diso=0$) in MBPT up to fourth order, including both NN and 3N interactions at the normal-ordered two-body level~\cite{Hebeler:2020ocj,Drischler:2021kxf}.
These MBPT calculations were carried out for the six Hebeler et~al.\ interactions, which are commonly used in nuclear structure and infinite matter calculations (see, e.g., Refs.~\cite{Simonis:2017dny,Drischler:2015eba,Drischler:2017wtt,Stroberg2021a,Miyagi:2021pdc,Arthuis:2024mnl,Bonaiti2025}).
They were obtained in Ref.~\cite{Hebeler:2010xb} by softening the \NNNLO NN potential EM~$500\MeV$ using the similarity renormalization group (SRG) and then adding them at different SRG resolution scales $\lambda_\text{SRG}$ to the unevolved \NNLO 3N forces.
The two 3N LECs at this chiral order, $c_D$ and $c_E$, were adjusted to the \isotope[3]{H} binding energy and the \isotope[4]{He} charge radius for two 3N momentum cutoffs $\Lambda_{\mathrm{3N}}$.
These interactions are labeled as ``$(\lambda_\text{SRG}, \Lambda_{\mathrm{3N}})$.''

The MBPT training data are provided on a regular density grid, $n = 0.01, 0.02, \dots, 0.32 \fmiq$, and a regular isospin-asymmetry grid, $\diso = 0, 0.1, \dots, 1.0$, yielding $32 \times 11 = 352$ data points for each interaction. 
Each training point comes with an uncorrelated estimate of the statistical uncertainty from the MC integrations of the MBPT diagrams performed in Ref.~\cite{Drischler:2026vdm}.
Exploiting the approximate symmetry of the EOS under neutron-proton exchange, $E(\diso,n)/A \approx E(-\diso,n)/A$, we augment the dataset by reflecting the EOS about $\diso=0$. 
The resulting grid spans $\diso=-1.0,-0.9,\ldots,0.9,1.0$ and contains $N=32\times 21=672$ training points. 
Using the expected (approximate) isospin symmetry of the EOS, this preprocessing step mitigates GP boundary effects by shifting the region around SNM from the edge of the training domain into the interpolation region.

Figure~\ref{fig:anm_mbpt} shows the results of these MBPT calculations (see also Figure~7 in Ref.~\cite{Drischler:2026vdm}).
Each panel depicts the energy per particle as a function of the density $n$ for a given isospin asymmetry $\diso$, as annotated.
The black dots depict the sample mean values and the error bars depict twice the standard deviations across the EOS calculations based on these six interactions.
We will discuss the blue bands when we revisit Fig.~\ref{fig:anm_mbpt} in Sec.~\ref {sec:discrepancy_model}.
In addition to the sample mean and standard deviation of the energy per nucleon on the $(\diso, n)$ grid, we quantify the correlations between different grid points. 
These correlations are estimated from the MBPT calculations through the unbiased sample covariance matrix (see Sec.~\ref{sec:deviation_kernel}).

\subsection{Statistical model}
\label{sec:discrepancy_model}

As discussed in Sec.~\ref{sec:micro_data}, we consider an ensemble of microscopic EOS calculations
$\{y^{(h)}(\vb{x}_t)\}_{h=1}^H$ at the common training inputs $\vb{x}_t$ obtained from the $H=6$ Hebeler et al.\ interactions.
Here, $\vb{y}_t^{(h)} \equiv y^{(h)}(\vb{x}_t)$ denotes the vector of computed energies per particle $E(\vb{x})/A$ associated with the interaction $h$, where each training input is
$\vb{x}_{t,i}=(\delta_{\mathrm{iso},i},n_i)$.
Because these EOS calculations are performed within the same theoretical framework, namely MBPT applied to a set of similar interactions based on the same calibration protocol~\cite{Hebeler:2010xb}, we model them as noisy realizations of a shared underlying mean EOS process, with interaction-specific deviations that are correlated in the input domain.\footnote{%
We assess our model assumptions in Sec.~\ref{sec:results} by comparing our results with those obtained from deterministic methods.%
} 
Accordingly, we assume a hierarchical model\footnote{%
In this section, the interaction-dependent deviation term $\delta(\vb{x})$ should not be confused with the Dirac delta $\delta_{\vb{x},\vb{x}'}$ or the isospin asymmetry $\diso$.%
} for this EOS process ($h = 1, 2, \ldots$):
\begin{equation} \label{eq:discrepancy_model}
Y^{(h)}(\vb{x}) = \eta(\vb{x}) + \delta^{(h)}(\vb{x}) + \varepsilon^{(h)}(\vb{x}) \,,
\end{equation}
where $\delta^{(h)}(\vb{x})$ represents the interaction-dependent deviation from the common mean and $\varepsilon^{(h)}(\vb{x})$ accounts for numerical noise in the microscopic EOS calculations.
The spread among the six computed EOSs is therefore not treated as numerical noise in the GP inference, but rather represents a region of unobserved EOSs.
We refer to $\delta^{(h)}(\vb{x})$ as a deviation rather than a model discrepancy, since it does not represent a bias with respect to the unobserved true EOS.
Following standard statistical notation, we use uppercase $Y^{(h)}(\vb{x})$ to denote the EOS stochastic process and lowercase $y^{(h)}(\vb{x})$ for its observed realization.

The central object in the hierarchical model~\eqref{eq:discrepancy_model} is the latent common mean EOS, $\eta(\vb{x})$, which represents the systematic behavior shared across the ensemble.
We place a GP prior on that common mean EOS,
\begin{equation} \label{eq:gp_prior_eta}
\eta \sim \mathcal{GP}\left( \mu_\eta, k_\eta \right),
\end{equation}
and model the interaction-dependent deviations as independent instances of a single underlying GP,\footnote{
In statistics, ``i.i.d.'' stands for independent and identically distributed; i.e., while a collection of random variables shares the same probability distribution, they are mutually independent (and exchangeable).
}
\begin{equation} \label{eq:gp_delta_a}
\delta^{(h)} \stackrel{\mathrm{i.i.d.}}{\sim} \mathcal{GP}\left (0, k_\delta \right),
\qquad h = 1, 2, \dots .
\end{equation}
Equation~\eqref{eq:gp_delta_a} reflects the assumption that the interactions are constructed within the same theoretical framework and evaluated on the same grid, and are therefore treated as exchangeable (under the prior assumption) with respect to their deviation structure.
The deviation GP is taken to have zero mean so that any systematic trend across interactions is absorbed by $\eta(\vb{x})$.
We further assume independent heteroscedastic white noise, 
\begin{equation}
\varepsilon^{(h)} \stackrel{\mathrm{i.i.d.}}{\sim} \mathcal{GP}\left(0, \sigma_\varepsilon^2(\vb{x}) \delta_{\vb{x}, \vb{x}'} \right),
\end{equation}
where $\sigma_\varepsilon^2(\vb{x})$ denotes an input-dependent variance, capturing the heteroscedastic numerical noise across the input space, and $\delta_{\vb{x}, \vb{x}'}$ is the Kronecker delta representing the white noise. 
The explicit forms of the mean function $\mu_\eta$ and the kernels $k_\eta$ and $k_\delta$ are specified in Secs.~\ref{sec:results} (and~\ref{sec:deviation_kernel}).

Our objective is not only to infer the common mean EOS, but to predict a noise-free EOS associated with a previously unobserved interaction drawn from the same population.
We denote this prediction by
\begin{equation}
\label{eq:latent_eos}
f^{\mathrm{(new)}}(\vb{x})
=
\eta(\vb{x})
+
\delta^{\mathrm{(new)}}(\vb{x}) \,,
\end{equation}
where $\delta^{\mathrm{(new)}}$ is an independent draw from the deviation GP in Eq.~\eqref{eq:gp_delta_a} and is independent of both $\eta$ and the deviations associated with the observed interactions.
The process $f^{\mathrm{(new)}}$ is noise-free, as it represents the latent EOS for a new interaction rather than the outcome of a noisy microscopic calculation.

Constructing the posterior for $f^{\mathrm{(new)}}$ requires both the covariance $k_\delta$ of the interaction-dependent deviations and the covariance $k_\eta$ of the common mean EOS.
We therefore calibrate the model in two stages.
First, we infer $k_\delta$ from the residuals of the individual EOSs about their sample mean.
Second, holding $k_\delta$ fixed, we infer $k_\eta$ from the ensemble mean while accounting for the finite number of interactions and the numerical noise.
Finally, we combine the posterior uncertainty in the common mean EOS with the full interaction-dependent covariance to obtain predictions for a new noise-free EOS.

\subsubsection{Covariance structure of the EOS ensemble}
\label{sec:hierarchical_covariance}

The cross-covariance implied by Eqs.~\eqref{eq:discrepancy_model}--\eqref{eq:gp_delta_a} is
\begin{multline}
\cov\left(
Y^{(h)}(\vb{x}),
Y^{(h')}(\vb{x}')
\right)
=
k_\eta(\vb{x},\vb{x}')
\\
+
\delta_{h,h'}
\left[
k_\delta(\vb{x},\vb{x}')
+
\sigma_\varepsilon^2(\vb{x})
\delta_{\vb{x},\vb{x}'}
\right] \,,
\end{multline}
where $\delta_{h,h'}$ is the Kronecker delta in interaction space.
Thus, the common mean process $\eta(\vb{x})$ correlates all EOS processes, whereas
$\delta^{(h)}(\vb{x})$ and $\varepsilon^{(h)}(\vb{x})$ contribute only when $h=h'$.
These EOS processes are conditionally independent given $\eta$, but they are correlated after marginalizing over the shared mean process due to the uncertainty in the mean EOS.
In practice, the latent mean process $\eta(\vb{x})$ is not directly observed, and inference is therefore based on its marginal effect on the EOS.
Marginalizing over $\eta$ yields that each EOS is itself a GP,
\begin{equation}
    Y^{(h)}(\vb{x}) \sim \mathcal{GP} \left(\mu_\eta(\vb{x}), k_{\eta}(\vb{x}, \vb{x}') + k_{\delta}(\vb{x}, \vb{x}') + \sigma_\varepsilon^2(\vb{x}) \delta_{\vb{x}, \vb{x}'} \right).
\label{eq:Yh-GP}
\end{equation}

To write the corresponding discretized model, let $\{\vb{x}_i\}_{i=1}^N$ denote the common set of $N$ training points introduced in Sec.~\ref{sec:micro_data}. 
For the interaction indexed by $h$, the $E/A$ random variable is represented by the length-$N$ vector
\begin{equation}
\vb{Y}^{(h)} = \vb*{\eta} + \vb*{\delta}^{(h)} + \vb*{\varepsilon}^{(h)} \,.
\end{equation}
Conditioned on the latent mean vector $\vb*{\eta}$, each discretized EOS then follows
\begin{equation}
\label{eq:marg_likelihood}
\vb{Y}^{(h)} \given \vb*{\eta}
\sim
\mathcal{N}\left(
\vb*{\eta},
\vb{K}_\delta + \vb{\Sigma}_{\text{MC}}
\right),
\end{equation}
where $\vb{K}_\delta$ is the covariance matrix generated by the kernel $k_\delta$ on the training grid. 
The diagonal covariance matrix $\vb{\Sigma}_{\text{MC}} = \text{diag}(\vb*{\sigma}_{\text{MC}}^2)$ accounts for the input-dependent numerical noise in the MC momentum integrations in Ref.~\cite{Drischler:2026vdm}, where $\vb*{\sigma}_{\mathrm{MC}}$ contains the average of the reported standard deviations across the six interactions estimated by the MBPT calculations at the grid points.

Since $\vb*{\eta}$ is not directly observed in practice, we marginalize over its GP prior by carrying out the integral
\begin{equation} \label{eq:eta_marginalization}
\pr\left(\vb{Y}^{(h)}\right)
=
\int
\pr\left(\vb{Y}^{(h)} \given \vb*{\eta}\right)
\, \pr\left(\vb*{\eta}\right)
\, \dd \vb*{\eta}\,.
\end{equation}
The integral in Eq.~\eqref{eq:eta_marginalization} can be done analytically because both the conditional likelihood~\eqref{eq:marg_likelihood} and the prior~\eqref{eq:gp_prior_eta} are Gaussian, resulting in a Gaussian marginal distribution:
\begin{equation}
\vb{Y}^{(h)}
\sim
\mathcal{N}\left(
\vb*{\mu}_{\eta},
\vb{K}_{\eta} + \vb{K}_{\delta} + \vb{\Sigma}_{\text{MC}}
\right).
\end{equation}
Likewise, for arbitrary evaluation points $\vb{x}$, each EOS realization follows the same GP as given in Eq.~\eqref{eq:Yh-GP}.

\subsubsection{Distribution of the ensemble mean}

To determine how the common mean EOS is informed by the ensemble, we first derive the distribution of the ensemble average. 
Let us define the stochastic ensemble mean
\begin{equation} \label{eq:ens_mean}
\bar{Y}(\vb{x}) = \frac{1}{H}\sum_{h=1}^H Y^{(h)}(\vb{x}) \,.
\end{equation}
Using the statistical model~\eqref{eq:discrepancy_model}, the ensemble mean~\eqref{eq:ens_mean} can be written as
\begin{equation}
\bar{Y}(\vb{x}) = \eta(\vb{x}) + \bar{\delta}(\vb{x}) + \bar{\varepsilon}(\vb{x}) \,,
\end{equation}
where $\bar{\delta}(\vb{x})$ and $\bar{\varepsilon}(\vb{x})$ are the corresponding mean values over interactions, defined similarly to Eq.~\eqref{eq:ens_mean}.
Because the interaction-dependent deviations $\delta^{(h)}(\vb{x})$ are assumed to be independent draws from a common zero-mean GP, their average is also a GP,
\begin{equation}
\bar{\delta} \sim \mathcal{GP}\left( 0,\frac{1}{H}k_\delta \right),
\end{equation}
and the mean numerical noise is also white,
\begin{equation}
\bar{\varepsilon} \sim \mathcal{GP}\left(0,\frac{\sigma_\varepsilon^2(\vb{x})}{H}\delta_{\vb{x},\vb{x}'}\right).
\end{equation}
Therefore, conditional on the common mean process $\eta(\vb{x})$, the ensemble mean~\eqref{eq:ens_mean} is then distributed as
\begin{equation}
\bar{Y}(\vb{x}) \given \eta (\vb{x})
\sim
\mathcal{GP}\left(
\eta(\vb{x}),
\frac{1}{H}k_\delta(\vb{x},\vb{x}')
+
\frac{\sigma_\varepsilon^2 (\vb{x})}{H}\delta_{\vb{x},\vb{x}'}
\right).
\end{equation}
Averaging over the $H$ interactions thus reduces both the interaction-dependent uncertainty and the numerical noise by the usual factor of $1/H$, while leaving the common mean process $\eta$ unchanged.
This reduction is what allows the ensemble average to inform $\eta$ without treating the interaction-to-interaction spread in the EOS as ordinary observational noise.

Similar to Eq.~\eqref{eq:eta_marginalization}, marginalizing over the GP prior for $\eta$, the ensemble mean is also a GP,
\begin{equation}
\bar{Y}(\vb{x})
\sim
\mathcal{GP}\left(
\mu_\eta(\vb{x}),
k_\eta(\vb{x},\vb{x}')
+
\frac{1}{H}k_\delta(\vb{x},\vb{x}')
+
\frac{\sigma_\varepsilon^2 (\vb{x})}{H}\delta_{\vb{x},\vb{x}'}
\right).
\end{equation}
Hence, the observed sample mean $\bar{\vb{y}}(\vb{x}_t)=\bar{\vb{y}}_t$ can be seen as a noisy observation of $\vb*{\eta}(\vb{x}_t)$ at the training points, with associated prior covariance matrix
\begin{equation} \label{eq:K_ybar_tt}
\vb{K}_{\bar{y},tt}
=
\vb{K}_{\eta,tt}
+
\frac{1}{H}\vb{K}_{\delta,tt}
+
\frac{1}{H}\vb{\Sigma}_{\text{MC}}.
\end{equation}
Equivalently, the kernel of the ensemble mean is
\begin{equation} \label{eq:k_ybar}
k_{\bar{y}}(\vb{x},\vb{x}')
=
k_\eta(\vb{x},\vb{x}')
+
\frac{1}{H}k_\delta(\vb{x},\vb{x}')
+
\frac{1}{H} \sigma_\varepsilon^2 (\vb{x}) \delta_{\vb{x},\vb{x}'}. 
\end{equation}
The numerical-noise term is needed only at the observed points, where $\sigma_\varepsilon (\vb{x})$ is known. 
As $H \to \infty$, the contributions proportional to $1/H$ vanish and $\vb{K}_{\bar{y},tt}$ approaches $\vb{K}_{\eta,tt}$, provided this limit exists. 

The preceding result shows that the ensemble mean can be treated as an observation of $\eta$, but with an effective covariance containing the contribution $k_\delta/H$. 
Consequently, the posterior of the common mean EOS cannot be constructed until the covariance of the interaction-dependent deviations has been specified. 
We therefore turn next to the construction and calibration of $k_\delta$.

\subsubsection{Construction and calibration of the deviation kernel}
\label{sec:deviation_kernel}

The distribution derived above identifies $k_\delta$ as a required input for inference of the common mean EOS. The first calibration stage therefore determines the covariance of the interaction-dependent deviations. 
At the training points $\{\vb{x}_i\}_{i=1}^N$, we define the unbiased ensemble covariance matrix\footnote{%
Note that the covariance matrix and the scatter matrix are related via $\vb*{\Sigma}_\delta = \vb{S}_\delta / (H-1)$.%
}
corresponding to the ensemble mean~\eqref{eq:ens_mean}:
\begin{subequations} \label{eq:sample_cov_delta}
\begin{align}
\widetilde{\vb*{\Sigma}}_\delta
&=
\frac{1}{H-1}\sum_{h=1}^H
\vb{R}_t^{(h)} \vb{R}_t^{(h)\intercal}\,,\\
\vb{R}_t^{(h)}
&=
\vb{Y}_t^{(h)}-\bar{\vb{Y}}_t \,.
\end{align}
\end{subequations}
According to the statistical model~\eqref{eq:discrepancy_model}, $\vb*{\eta}_t$ cancels in the centered residuals $\vb{R}_t$ and thus they are Gaussian variables with zero mean and covariance matrix $((H-1)/H) \, (\vb{K}_{\delta,tt} + \vb{\Sigma}_{\text{MC}})$, resulting in
\begin{equation} \label{eq:wishart_S}
(H-1)\widetilde{\vb*{\Sigma}}_\delta \sim \mathcal{W}_N \left( \vb{K}_{\delta,tt} + \vb{\Sigma}_{\text{MC}},\,H-1 \right),
\end{equation}
where $\mathcal{W}_N$ denotes the $N$-dimensional Wishart distribution with $H-1$ degrees of freedom. 
We treat the (observed) unbiased sample covariance matrix $\vb*{\Sigma}_\delta$ obtained from the MBPT calculations as a realization of the random variable $\widetilde{\vb*{\Sigma}}_\delta$.
However, with only $H=6$ interactions, $\vb*{\Sigma}_\delta$ is necessarily singular and low-rank. 
Specifically, $\operatorname{rank}(\vb*{\Sigma}_\delta) \leqslant \min(N,H-1) $, so that $\vb*{\Sigma}_\delta$ has here rank of (at most) 5. 
Because of the significant rank deficiency of the observed $\vb*{\Sigma}_\delta$, we augment the deviation kernel by writing it as
\begin{multline} \label{eq:kdelta_sum}
k_\delta(\vb{x},\vb{x}';\vb*{\theta}_\delta)
=
\alpha_{\mathrm{emp}}\,k_\delta^{\mathrm{(emp)}}(\vb{x},\vb{x}')
\\ +
k_\delta^{\mathrm{(sm)}}(\vb{x},\vb{x}';\vb*{\theta}_{\mathrm{sm}}),
\end{multline}
with the scale parameter $\alpha_{\mathrm{emp}} \geqslant 0$,
and the joint set of hyperparameters $\vb*{\theta}_\delta = \{\alpha_{\mathrm{emp}},\vb*{\theta}_{\mathrm{sm}}\}$.
The first term in Eq.~\eqref{eq:kdelta_sum} is a low-rank empirical kernel (``emp'') constructed from the (dominant) eigenmodes of the centered sample covariance.
The second term in Eq.~\eqref{eq:kdelta_sum} is assumed to be a smooth, stationary kernel (``sm'') that regularizes the covariance and stabilizes interpolation away from the training points. 
For the residual kernel in Eq.~\eqref{eq:kdelta_sum}, we use an RBF kernel $k_\delta^{\mathrm{(sm)}}(\vb{x}, \vb{x}')$ with hyperparameters $\vb*{\theta}_{\mathrm{sm}}$. 
It provides a full-rank background covariance that accounts for interaction-dependent variation not fully captured by the small number of empirical eigenmodes.
Because both kernels are valid kernels, so is their sum, and covariance matrices generated from this sum kernel are positive definite.

The empirical kernel in Eq.~\eqref{eq:kdelta_sum} is obtained from the eigendecomposition, $\vb*{\Sigma}_\delta = \vb{U} \vb{\Lambda} \vb{U}^\intercal$, and the associated decomposition
\begin{equation} \label{eq:kdelta_empirical}
k_\delta^{\mathrm{(emp)}}(\vb{x},\vb{x}')
\simeq 
\sum_{m=1}^{M} \lambda_m \,\phi_m(\vb{x})\,\phi_m(\vb{x}') \,,
\end{equation}
with $M \leqslant H-1$, the eigenvalues and interpolated eigenfunctions, respectively corresponding to the eigenvalues and discrete eigenvectors of $\vb*{\Sigma}_\delta$, $\lambda_m$ and $\phi_m(\vb{x})$. 
Here, we consider the 3 dominant eigenmodes ($M=3$), so that Eq.~\eqref{eq:kdelta_empirical} is a low-rank approximation of $\vb*{\Sigma}_\delta$. 
These eigenmodes are interpolated with a GP via \gpdiff to obtain continuous representations.
We observe much higher numerical noise in the 2 remaining eigenvectors (associated with nonzero eigenvalues) and thus omit them, which is further justified by the amount of preserved variance $\eta_{\mathrm{p.v.}}$~\cite{Maldonado:2025ftg} being very close to unity, i.e., $1-\eta_{\mathrm{p.v.}} \approx 10^{-4}$.

The kernel hyperparameters $\vb*{\theta}_\delta$ are estimated by maximizing the restricted log-likelihood constructed from the original ensemble data~\cite{PattersonThompson1971,Harville1977}. 
For the present Gaussian model, formally integrating out the likelihood over the unknown mean vector under an improper flat prior results in the restricted likelihood~\cite{Harville1974}.
Up to additive constants independent of $\vb*{\theta}_\delta$, the restricted log-likelihood reads [corresponding to Eq.~\eqref{eq:wishart_S}]
\begin{equation} \label{eq:K_delta_likelihood}
\mathcal{L}(\vb*{\theta}_\delta)
=
-\frac{H-1}{2}\ln \left| \vb{K}_{tt} \right|
-\frac{H-1}{2}
\tr \left[ \vb{K}_{tt}^{-1}\vb{\Sigma}_\delta \right],
\end{equation}
where $\vb{K}_{tt} = \vb{K}_{\delta, tt} + \vb{\Sigma}_{\text{MC}}$ and $\vb{K}_{\delta, tt}$ is generated by the kernel~\eqref{eq:kdelta_sum}. 
The trace term in Eq.~\eqref{eq:K_delta_likelihood} enforces agreement with $\vb{\Sigma}_\delta$, while the log term penalizes overly complex kernels. 
Note that the inverse of the rank-deficient $\vb{\Sigma}_\delta$ does not enter the likelihood~\eqref{eq:K_delta_likelihood}.
To improve numerical stability, both terms are evaluated using the Cholesky decomposition.
One can show~\cite{lee2026thesis} that at the maximum of the restricted log-likelihood~\eqref{eq:K_delta_likelihood} with respect to arbitrary kernel variations and full-rank $\vb*{\Sigma}_{\delta}$ is $\vb{K}_{tt} = \vb*{\Sigma}_{\delta}$, as required.
However, in our case, $\vb*{\Sigma}_{\delta}$ does not have full rank and $k_\delta$ is restricted to the chosen form in Eq.~\eqref{eq:kdelta_sum}, so $\vb{K}_{tt} \approx \vb*{\Sigma}_{\delta}$.

\subsubsection{Posterior of the common mean EOS}
\label{sec:eos_mean}

With the deviation kernel now calibrated, all components of the ensemble mean covariance are specified. We can therefore determine the common mean process by conditioning on the observed sample mean. 
At arbitrary evaluation points,
\begin{equation}
\vb*{\eta} \given \bar{\vb{y}}_t, \vb{x}_t, \vb{x}_*
\sim
\mathcal{N}\left(\vb{m}_{\eta,*|t}, \vb*{K}_{\eta,**|t}\right),
\end{equation}
with posterior mean and posterior covariance matrix, respectively, given by
\begin{align}
\vb{m}_{\eta,*|t}
&=
\vb*{\mu}_{\eta,*}
+
\vb{K}_{\eta,*t}
\vb{K}_{\bar{y},tt}^{-1}
\left(\bar{\vb{y}}_t-\vb*{\mu}_{\eta,t} \right),
\label{eq:eta_posterior_mean}
\\
\vb{K}_{\eta,**|t}
&=
\vb{K}_{\eta,**}
-
\vb{K}_{\eta,*t}
\vb{K}_{\bar{y},tt}^{-1}
\vb{K}_{\eta,t*} \,,
\label{eq:eta_posterior_cov}
\end{align}
where the cross-covariance is given by
\begin{equation}
    [\vb{K}_{\eta,*t}]_{ij} = \cov ( \eta(\vb{x}_{*i}),\bar Y(\vb{x}_{t,j}) ) = k_\eta (\vb{x}_{*i},\vb{x}_{t,j})
\end{equation}
because $\bar{\delta}(\vb{x})$ and $\bar{\varepsilon}(\vb{x})$ are independent of $\eta(\vb{x})$.
While the posterior mean~\eqref{eq:eta_posterior_mean} provides the best estimate of the common EOS shared across the ensemble, $\vb*{\eta}$, the posterior covariance~\eqref{eq:eta_posterior_cov} quantifies the remaining uncertainty in $\vb*{\eta}$ after accounting for the finite sample size and remaining numerical noise of the ensemble average. 

To complete the specification of the statistical model, the kernels $k_\eta$ and $k_\delta$, along with their hyperparameters, must be defined and calibrated to the considered EOS ensemble. 
Because these kernels govern distinct aspects of the statistical model~\eqref{eq:discrepancy_model}, we adopt a two-step calibration procedure. 
First, the hyperparameters of $k_\delta (\vb*{\theta}_\delta$) are calibrated to the observed (unbiased) ensemble covariance matrix $\Sigma_{\delta}$, which isolates the interaction-dependent deviation about the common mean (see Sec.~\ref{sec:deviation_kernel} for details). 
Second, given $k_\delta$, the hyperparameters of $k_\eta (\vb*{\theta}_\eta$), governing the prior covariance structure of $\eta(x)$, are calibrated to the sample mean $\bar{\vb{y}}_t$ with associated uncertainties encoded in Eq.~\eqref{eq:K_ybar_tt}.
This construction ensures that the inference for $k_\eta$ explicitly accounts for the (reduced) interaction-dependent deviation (of the finite EOS sample size) and the reduced numerical noise already present in $\bar{\vb{y}}_t$, thereby avoiding double counting of uncertainty between $\eta$ and $\delta$.

We infer the hyperparameters $\vb*{\theta}_\eta$ by maximizing the Gaussian log-marginal likelihood, which reads up to additive constants independent of $\vb*{\theta}_\eta$:
\begin{equation}
\ln \pr (\bar{\vb{y}}_t \given \vb x_t,\vb{K}_{\delta,tt},\vb*{\theta}_\eta)
=
-\frac{1}{2}\tilde{\vb{y}}_t^\intercal \vb{K}_{\bar y,tt}^{-1}\tilde{\vb{y}}_t
-\frac{1}{2}\ln |\vb{K}_{\bar y,tt}|\,,
\end{equation}
with $\tilde{\vb{y}}_t=\bar{\vb{y}}_t-\vb*{\mu}_{\eta,t}$. In this step, $\vb{K}_{\delta,tt}$ is treated as fixed at its estimated value obtained in Sec.~\ref{sec:deviation_kernel}. 

\subsubsection{Predictive distribution for a new EOS}
\label{sec:model_output}

We now return to the predictive target in Eq.~\eqref{eq:latent_eos}.
Before conditioning on the MBPT data for the EOS, $f^{\mathrm{(new)}}(\vb{x})$ is marginally distributed as
\begin{equation}
f^{\mathrm{(new)}}(\vb{x})
\sim
\mathcal{GP}\left(
\mu_\eta(\vb{x}),
k_\eta(\vb{x},\vb{x}')
+
k_\delta(\vb{x},\vb{x}')
\right) \,,
\end{equation}
corresponding to Eq.~\eqref{eq:eta_marginalization} when also the numerical noise is marginalized over and with the kernels defined in Eqs.~\eqref{eq:k_ybar} and~\eqref{eq:kdelta_sum}.
Hence, at evaluation points $\vb{x}_*$ one has the distribution:
\begin{subequations} \label{eq:eos_posterior}
\begin{align}
f^{\mathrm{(new)}}(\vb{x}_*) \given \bar{\vb{y}}_t
&\sim
\mathcal{N}\left(\vb{m}_{f,*|t}, \vb{K}_{f,**|t} \right) \,, \\
\vb{m}_{f,*|t} &= \vb{m}_{\eta,*|t} \,, \\ 
\vb{K}_{f,**|t}
&=
\vb{K}_{\eta,**|t}
+
\vb{K}_{\delta,**} \,. \label{eq:sigma_fstar}
\end{align}
\end{subequations}
By construction, the predictive mean $\vb{m}_{f,*|t}$ is equal to that of $\vb{m}_{\eta,*|t}$ defined in Eq.~\eqref{eq:eta_posterior_mean}, whereas the predictive covariance $\vb{K}_{f,**|t}$ obtains a contribution from the interaction-dependent deviation in addition to $\vb{K}_{\eta,**|t}$ defined in Eq.~\eqref{eq:eta_posterior_cov}.
Both can be calculated at arbitrary evaluation points.

Equation~\eqref{eq:eos_posterior} shows that, if $H \ll \infty$, the uncertainty bands for these EOS predictions are naturally wider than those of the latent mean EOS and the interaction-dependent deviation encoded in the sample covariance matrix, $\vb*{\Sigma}_\delta$, individually.
One can see this feature also in Fig.~\ref{fig:anm_mbpt}, where the blue bands depicting our GP model at the $2\sigma$ confidence level are slightly larger than the black error bars on which it was trained. 
This larger uncertainty is because $\vb*{\Sigma}_\delta$ is defined in terms of the residuals $\{\vb{y}_t^{(h)} - \bar{\vb{y}}_t\}_{h=1}^H$ [see Eq.~\eqref{eq:sample_cov_delta}], causing the common mean $\eta(\vb{x})$ in our statistical model~\eqref{eq:discrepancy_model} to cancel. 
As a result, $\vb*{\Sigma}_\delta$ captures only the interaction-dependent uncertainty about the ensemble mean, whereas the predictive covariance for a single noise-free EOS prediction, which we are mainly interested in, should also capture the uncertainty in the predicted mean.
On the other hand, the numerical noise contributes only through the likelihood functions and is excluded from the final predictive distribution of the noise-free EOS.

\begin{figure*}[tb]
    \includegraphics[width=\textwidth]{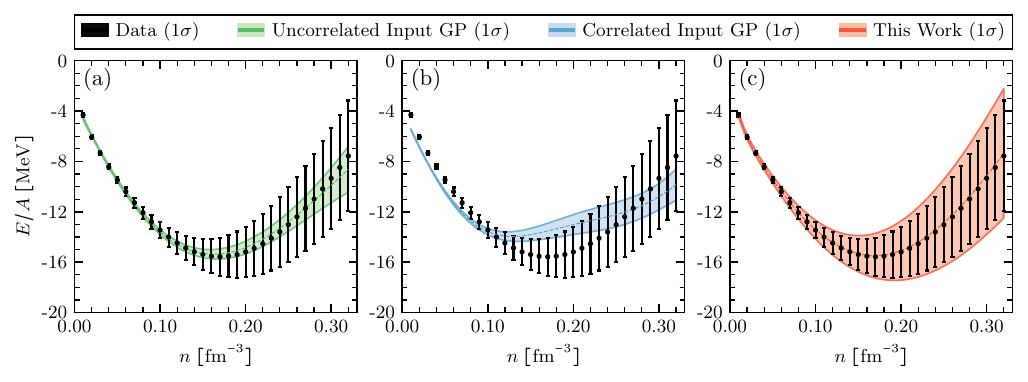} 
    \caption{%
    Comparison of the one-dimensional GP regression based on different model assumptions (for illustration).
    Panels~(a) and~(b) show straightforward GP regressions with the diagonal $\vb{C}_t = \diag(\vb{\Sigma}_{\delta} + \vb{\Sigma}_{\text{MC}})$ and fully occupied $\vb{C}_t = \vb{\Sigma}_{\delta} + \vb{\Sigma}_{\text{MC}}$ treated as the numerical noise term for training, respectively. 
    Panel~(c) shows the GP regression with the proposed statistical model with the mean kernel and the deviation kernel. 
    This illustration is based on the SNM data shown in Fig.~\ref{fig:anm_mbpt}. 
    The black center dots and vertical bars in each panel depict the empirical mean and $1\sigma$ credibility region of the MBPT calculations, and the colored dashed lines and bands depict the predicted mean and $1\sigma$ uncertainty interval from the GP regressions, respectively.%
    }
    \label{fig:comp_gp_models}
\end{figure*}

Figure~\ref{fig:comp_gp_models} compares the predictions of GPs (shaded bands) trained on the same dataset (shown as error bars) using \gpdiff and different model assumptions. 
For this illustration, we use the SNM data from Fig.~\ref{fig:anm_mbpt}, where the error bars represent the sample mean $\bar{\vb{y}}_t$ and covariance matrix $\vb*{\Sigma}_\delta$ evaluated at the training points, as discussed in Sec.~\ref{sec:micro_data}. 
Panel~(a) shows the predictions obtained when only the diagonal covariance matrix $\vb{C}_{t} = \diag(\vb*{\Sigma}_\delta + \vb{\Sigma}_{\text{MC}})$ is treated as the numerical noise term for standard GP training (thereby ignoring all correlations in the uncertainty estimates),\footnote{%
The scenario in panel~(a) could be straightforwardly realized in \texttt{scikit-learn} by setting the GP regressor's input parameter $\vb*{\alpha}$ to $\diag(\vb*{\Sigma}_\delta + \vb{\Sigma}_{\text{MC}})$.%
} 
while panel~(b) corresponds to training with the fully occupied $\vb{C}_{t} = \vb*{\Sigma}_\delta + \vb{\Sigma}_{\text{MC}}$. 
In both panels~(a) and~(b), the GP hyperparameters are optimized by maximizing the likelihood in Eq.~\eqref{eq:marg_like}.
These two panels show the general feature of GPs that, in these cases, the predicted uncertainties are smaller than those of the data, which is not desired in our EOS analysis.
In contrast to the behaviors observed in panels~(a) and~(b) of Fig.~\ref{fig:comp_gp_models}, for our EOS analysis, we aim to construct models that are statistically consistent with the empirical mean and covariance inferred from the many-body data, as shown in panel~(c).
This panel depicts the corresponding results obtained by modeling $\vb*{\Sigma}_\delta$ as the deviation kernel $\vb{K}_\delta$ based on the framework described in Sec.~\ref{sec:deviation_kernel}, which is then used to train the mean EOS kernel $\vb{K}_\eta$ in Sec.~\ref{sec:eos_mean} and to derive the EOS mean and uncertainty via Eq.~\eqref{eq:eos_posterior}.
The GP regression in panel~(c), based on our statistical model, achieves the desired consistency between the MBPT data and the GP prediction, including uncertainties, and is therefore adopted in this work.

\section{Results and discussion}
\label{sec:results}

In this section, we present our main results.
We use \gpdiff together with the predictive distribution~\eqref{eq:eos_posterior} for quantifying and propagating correlated EOS uncertainties to derived quantities.

To explore CS kernels and probe the sensitivity of our constraints to our kernel choices (within a specific subclass), we examine two cases: 
a single RBF kernel and a CS kernel composed of three RBF kernels in the mean-EOS GP.
For the latter kernel, the centers are placed at $\diso = 0, \pm 1$ and $n = 0.16 \fmiq$, allowing the kernel to distinguish different features in PNM and SNM.%
\footnote{%
The two RBF kernels at $\diso = \pm 1$ share the same hyperparameters to ensure charge symmetry, amounting to 6 free hyperparameters in total.%
}
Here, we choose fixed values $a_i = 1$ and $\vb{B}_i = \diag(0.5,0.5)$ for all $i$, in the respective units, to facilitate a smooth crossover between the kernels predominant in PNM and SNM, respectively.
The density length scale is arbitrary in this case, because the kernels are aligned at a fixed density.
In both cases, we use the zero-mean function.

We obtain qualitatively similar results for both kernel choices, and therefore focus in this section on the results obtained with the CS kernel. 
Our results for the other kernel choice can be found in Appendix~\ref{app:add_results}.
These findings suggest that a smooth RBF kernel with a single length scale in each dimension for the mean EOS is sufficiently flexible to capture the dominant features of low-density asymmetric matter.
However, this conclusion should be revisited when modeling the EOS over a wider density range, e.g., for applications to neutron stars~\cite{Semposki:2025etb}.

\subsection{Low-density EOS parameters}
\label{sec:constaints_eos_params}

\begin{figure*}[p]
        \includegraphics[width=\linewidth]{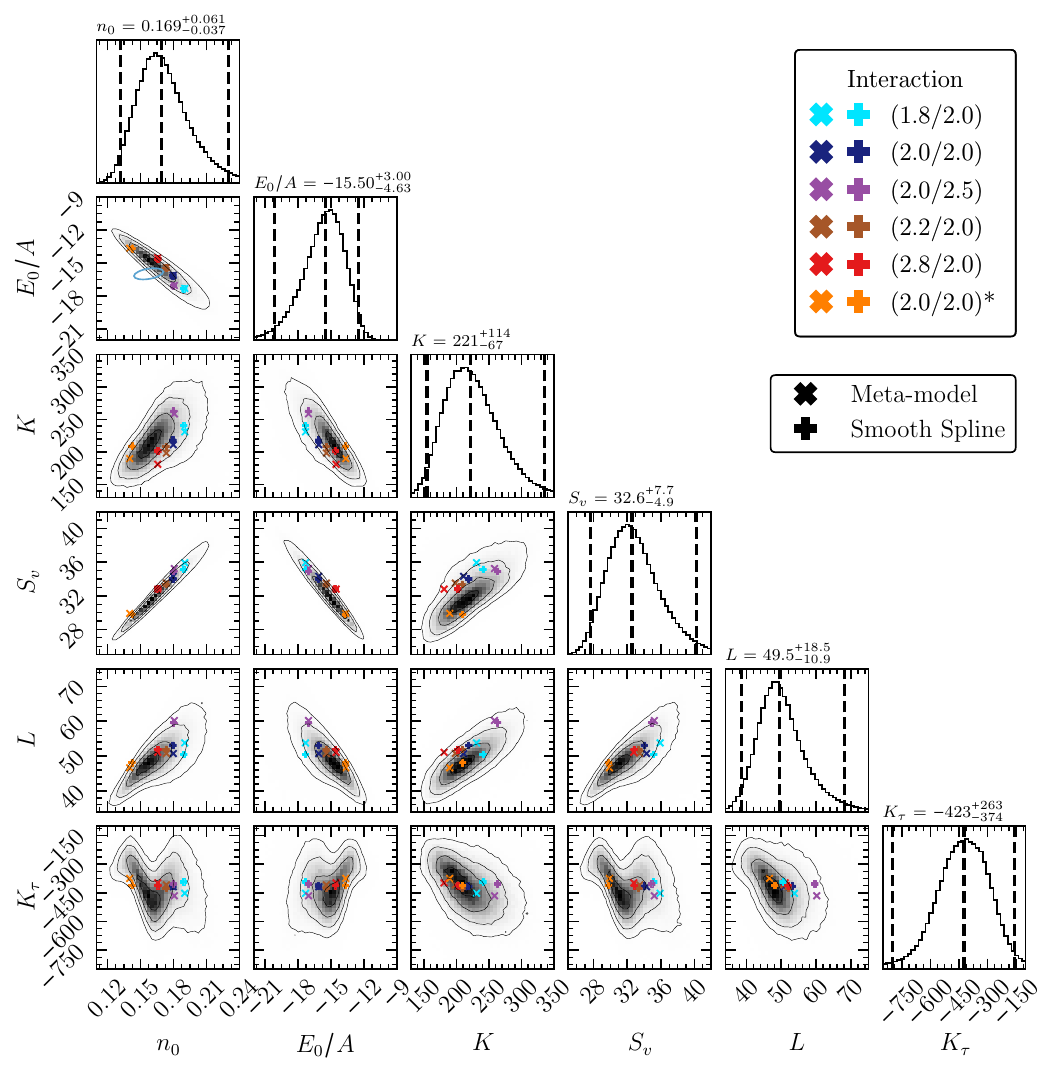}
        \caption{%
        Posterior distributions of the inferred low-density EOS parameters. 
        The parameters include the saturation density $n_0$, saturation energy $E_0/A$, incompressibility $K$, symmetry energy $S_v$, slope parameter $L$, and isospin incompressibility parameter $K_\tau$, all evaluated at $n_0$. 
        All parameters are given in $\MeV$, except for $n_0$, which is in $\fmiq$. 
        Diagonal panels show marginalized posterior distributions, with titles reporting the medians and central 95\% credibility intervals corresponding to the dashed lines. 
        Contours correspond to the approximate $0.5\sigma$, $1\sigma$, $1.5\sigma$, and $2\sigma$ confidence regions. 
        Symbols indicate the corresponding values for the six Hebeler et al.\ interactions extracted using the ``Meta-model''~($\bm{\times}$) and ``Smooth Spline'' method~($\bm{+}$), as described in the main text.
        The blue ellipse in the panel showing the $n_0$--$E_0$ correlations depicts the empirical saturation point at the 95\% credibility level determined in Ref.~\cite{Drischler:2024ebw}.%
        }
    \label{fig:corner_emp_eos_params}
\end{figure*}

We constrain the following parameters that describe the low-density, asymmetric matter EOS.
The SNM saturation density and energy are given by
\begin{subequations}
\begin{align}
    n_0 &\equiv n_0(\diso=0) \,,\\
    \frac{E_0}{A}&\equiv \frac{E}{A}(\diso,n)\bigg|_{\substack{n=n_0\\\diso = 0}} \label{eq:E0}\,.
\end{align}
\end{subequations}
We define the saturation density $n_0(\diso)$ as
\begin{equation} \label{eq:n0-delta}
    \frac{\partial}{\partial n}
    \frac{E}{A}(\diso,n)
    \bigg|_{n=n_0(\diso)}
    =0 \,.
\end{equation}
The leading isospin-dependent EOS parameters are the nuclear symmetry energy and its slope parameter,
\begin{align}
S_v  &\equiv \frac{1}{2} \frac{\partial^2}{\partial \diso^2} \frac{E}{A}(\diso,n) \bigg|_{\substack{n=n_0\\\diso = 0}}\,, \label{eq:Sv}\\ 
L  &\equiv \frac{3n_0}{2} \frac{\partial^3}{\partial n \partial \diso^2}  \frac{E}{A}(\diso,n) \bigg|_{\substack{n=n_0\\\diso = 0}}\,. \label{eq:L}
\end{align}
We also consider the isospin-dependent incompressibility of asymmetric matter,\footnote{%
See also Refs.~\cite{Piekarewicz:2008nh,Piekarewicz:2009gb} for detailed discussions of the incompressibility and related low-density EOS parameters.%
}
\begin{equation} \label{eq:K-delta}
   K(\diso) = 9 n_0^2(\diso) \frac{\partial^2}{\partial n^2}  \frac{E}{A}(\diso,n) \bigg|_{\substack{n=n_0(\diso)}}\,. 
\end{equation}
Expanding around SNM, one gets two leading coefficients of the series expansion of Eq.~\eqref{eq:K-delta},
\begin{equation} \label{eq:K-delta-exp}
    K(\diso) = K + \diso^2 K_\tau + \mathcal{O}(\diso^4) \,,
\end{equation}
which are the incompressibility in SNM,
\begin{equation} \label{eq:K}
   K \equiv 9 n_0^2 \frac{\partial^2}{\partial n^2}  \frac{E}{A}(\diso,n) \bigg|_{\substack{n=n_0\\\diso=0}}\,. 
\end{equation}
and the isospin-asymmetry term,
\begin{subequations} \label{eq:K_tau}
    \begin{align}
K_\tau &= K_{\mathrm{sym}} - 6L - \frac{Q_0}{K} L\,,\\
    K_{\text{sym}} &= \frac{9n_0^2}{2} \frac{\partial^4}{\partial \diso^2 \partial n^2} \frac{E}{A}(\diso,n) \bigg|_{\substack{n=n_0\\\diso = 0}} \,,\\
Q_0 &= 27 n_0^3  \frac{\partial^3}{\partial n^3} \frac{E}{A}(\diso,n) \bigg|_{\substack{n=n_0\\\diso = 0}}\,.
\end{align}
\end{subequations}

To constrain these low-density EOS parameters, we jointly sample $E(\diso,n)/A$ and the involved derivatives using \gpdiff.
The GP is sampled on a two-dimensional grid of $n \in [0.01, 0.28] \fmiq$ (28 points) and $\diso \in [0,0.7]$ (8 points). 
Since all quantities are drawn simultaneously from the same GP, the correlations between the EOS and its derivatives, and between different values of $n$ and $\diso$, are automatically included.
Once $n_0$ is determined by interpolating $\partial(E/A)/\partial n$, the EOS parameters are evaluated at that density using Eqs.~\eqref{eq:E0}, \eqref{eq:Sv},
\eqref{eq:L}, \eqref{eq:K}, and~\eqref{eq:K_tau}. 
For each jointly drawn sample of the GP posterior, local cubic splines are constructed separately for $E/A$ and the correlated derivatives (in the GP sample), and then evaluated at $n_0$. 
In the rare cases when a GP sample does not saturate or has more than one local minimum at any $\diso$, the entire sample is discarded.

Figure~\ref{fig:corner_emp_eos_params} shows the resulting posterior distribution of the EOS parameters, based on 399,163 out of 400,000 samples that went through the filtering process.
The titles on the diagonal panels state the 95\% confidence regions centered on the medians of the corresponding marginal distributions (dashed vertical lines).
The contour lines, moving outward, enclose the $0.5\sigma$, $1\sigma$, $1.5\sigma$, and $2\sigma$ confidence regions. %
All parameters in Fig.~\ref{fig:corner_emp_eos_params} are given in $\MeV$, except for $n_0$, which is in $\fmiq$.

We observe a strong Coester-band-like anticorrelation between $(n_0, E_0/A)$~\cite{Drischler:2015eba,Drischler:2017wtt}.\footnote{%
To quantify correlations and anticorrelations, we adopt the terminology of Ref.~\cite[see the discussion of Eq.~(26)]{Drischler:2024ebw}.
} 
In the normal approximation, we obtain the corresponding mean vector and the covariance matrix
%
\begin{equation} %
    \vb*{\mu} \equiv \begin{bmatrix}
    n_0\\
    E_0/A
    \end{bmatrix}
    \approx \begin{bmatrix}
    0.172 \\ -15.7
    \end{bmatrix} \,,  
    \vb*{\Sigma} \approx \mqty[0.025^2 & -0.212^2 \\ -0.212^2 & 1.95^2] \,,
\end{equation}
with a Pearson correlation coefficient of $\rho = -0.93$. 
Consistent with previous studies~\cite{Drischler:2015eba,Drischler:2017wtt,Drischler:2026vdm}, the predicted saturation point overlaps with the empirical saturation point, which is depicted by the light-blue ellipse at the 95\% credibility level in Fig.~\ref{fig:corner_emp_eos_params}, but shifted toward higher saturation densities and lower saturation energies.
The constraints on the empirical saturation point were obtained in Ref.~\cite{Drischler:2024ebw} based on Bayesian model mixing of various Skyrme and RMF constraints.
Furthermore, we observe a still strong correlation between $(S_v,L)$, with $\rho = 0.76$ and 
%
\begin{equation} %
    \vb*{\mu} \equiv \begin{bmatrix}
    S_v\\
    L
    \end{bmatrix}
    \approx \begin{bmatrix}
    32.9 \\ 50.4
    \end{bmatrix} \,, \quad 
    \vb*{\Sigma} \approx \mqty[3.22^2 & 4.26^2 \\ 4.26^2 & 7.41^2] \,.
\end{equation}
In addition, we find $(S_v,n_0)$ strongly correlated and, correspondingly, $(S_v,E_0/A)$ strongly anticorrelated.
%
%

Our constraints on the incompressibility parameters are less precise. 
At the 68\% credibility level, we find $K=221^{+48}_{-38}\MeV$ and $K_\tau=-423^{+133}_{-145}\MeV$.
Both results are, within their large uncertainties, consistent with empirical constraints inferred from isoscalar giant monopole resonance (ISGMR) measurements.
In particular, Ref.~\cite{Roca-Maza:2018ujj} reported
$K^\mathrm{(emp)} \approx 240 \pm 20 \MeV$ as the phenomenological range based on several analyses of closed-shell nuclei.
Moreover, ISGMR measurements of the even-$A$ isotopes $\isotope[112-124]{Sn}$ and $\isotope[106-116]{Cd}$ extracted
$K_\tau^\mathrm{(emp)}= -550 \pm 100 \MeV$ and
$K_\tau^\mathrm{(emp)}=-555 \pm 75 \MeV$, respectively~\cite{PhysRevLett.99.162503,Li:2010kfa,PATEL2012447}, and the comprehensive analysis in Ref.~\cite{Stone:2014wza} based on even-even \isotope{Sn} and \isotope{Cd} isotopes obtained $250 < K^\mathrm{(emp)} < 315 \MeV$ and $K_\tau = -(840 - 350) \MeV$.
Hence, both the empirical and microscopic constraints have significant uncertainties, although ours are somewhat larger.
We note, however, that comparisons with these empirical constraints should be made with caution, since they are inferred from finite-nucleus observables, whereas our calculations are directly performed in infinite nuclear matter~\cite{Stone:2014wza,Basu:2009vz,Bonaiti:2025euf}.
Nevertheless, both our results and the empirical constraints indicate that $K_\tau < 0$.

To cross-check our GP-based constraints and the underlying model assumptions [see Sec.~\ref{sec:discrepancy_model}], we determine these EOS parameters for each of the six interactions independently using two GP-free methods. 
In the first method, we consider the parametric model~\cite{Wen:2020nqs}
\begin{subequations} \label{eq:e-a-semiana}
    \begin{align}
        \begin{split}
        \frac{E_\text{M}}{A}(\diso,n) &= A_0(n) + A_2(n)\diso^2 \\
        & \quad +  \left[ A_{4}(n) + A_{4}^{\text{(log)}}(n) \, \ln |\diso| \right]
     \diso^{4} \,,
        \end{split} \\
     A_\mu(n) &= \sum \limits_{\nu= 2,3,4,5,6} C_{\mu\nu} \, \left[ \frac{n}{n_0^\star} \right]^{\frac{\nu}{3}} \,,
    \end{align}
\end{subequations}
from which derivatives can then be computed symbolically, e.g., with \texttt{SymPy}.
Here, $n_0^\star$ is the fiducial value for the saturation density, $n_0^\star = 0.16 \fmiq$, and $C_{\mu\nu}$ are the 20 model parameters determined by a least-squares fit to the EOS data, weighted by the inverse of the estimated MC errors.
Note that the model~\eqref{eq:e-a-semiana} is based on the standard expansion of the energy per particle in $\diso^2$, truncated and supplemented by the logarithmic term at fourth order. 
This term, which does not contribute to the EOS in the limits of PNM and SNM, was identified in Ref.~\cite{Kaiser:2015qia} and later confirmed numerically in Ref.~\cite{Wellenhofer:2016lnl}.
The density dependence of the coefficients $A_\mu(n)$ is modeled following Ref.~\cite{Drischler:2015eba}; see also Ref.~\cite{Drischler:2026vdm}.
We construct the model~\eqref{eq:e-a-semiana} with the goal of obtaining precise reference constraints for our GP-based approach, rather than studying nonanalytic terms in the isospin-asymmetry expansion. 
For a comprehensive analysis of such terms, we refer the reader to Ref.~\cite{Wen:2020nqs}.

In the second method, we construct a smoothing bivariate spline as implemented in \texttt{scipy}'s \texttt{Smooth\allowbreak{}Bivariate\allowbreak{}Spline}~\cite{2020SciPy-NMeth}. 
The MC uncertainties are used as weights in the sum of squared deviations, with the smoothing parameter set to the dataset size, $s=672$, and spline degrees $k_n=4$ and $k_{\diso}=3$.
This spline method produces a differentiable interpolant that suppresses numerical noise in the EOS calculations.

The symbols in Fig.~\ref{fig:corner_emp_eos_params} depict the results for each of the six EOSs obtained using these methods: 
``Meta-model''~($\bm{\times}$) and ``Smooth Spline''~($\bm{+}$). 
For each EOS, the two GP-free methods result in similar values, with noticeable deviations only in the constraints on $K_\tau$ and its correlations, shown in the bottom row of Fig.~\ref{fig:corner_emp_eos_params}. 
However, the actual level of consistency is difficult to assess without uncertainty estimates. 
Furthermore, these results are consistent with those obtained from our GP-based approach, generally falling within the $\approx 2\sigma$ credible regions or better [e.g., see $(n_0,E_0/A)$ in Fig.~\ref{fig:corner_emp_eos_params}]. 
This consistency gives us confidence in our statistical approach to constraining derived observables from the noisy microscopic EOS calculations considered here and highlights the key advantage of our GP-based method over the GP-free approaches: 
uncertainty estimates are naturally built in.

\begin{figure*}[tb]
    \centering
    \includegraphics[width=\linewidth]{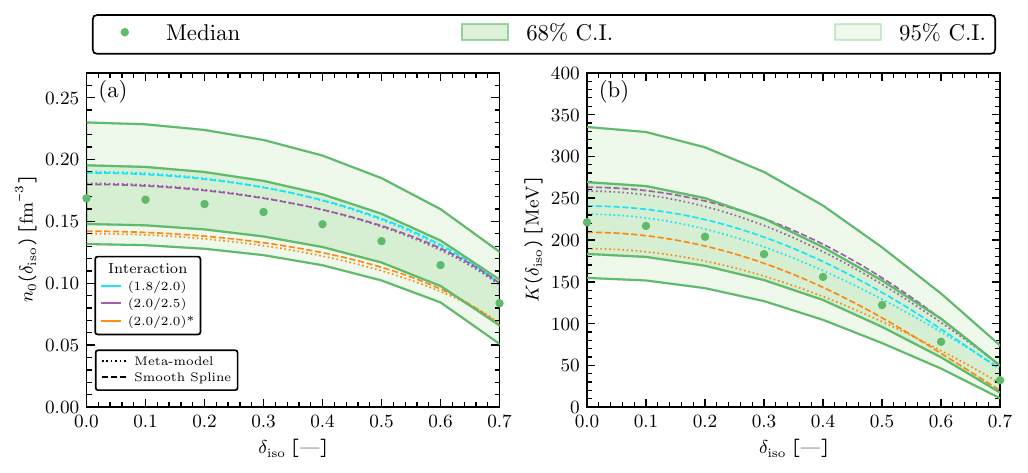}
    \caption{%
    The predicted saturation density, $n_0(\diso)$, and incompressibility, $K(\diso)$, as a function of the isospin asymmetry in panels~(a) and~(b), respectively. 
    The green dots represent the median, and the dark (light) shading corresponds to the 68\% (95\%) credibility interval~(C.I.).
    The other lines depict the results obtained from three representative individual EOS calculations using the ``Meta-model'' (dotted lines) and the ``Smooth Spline'' method (dashed lines).%
    }
    \label{fig:isospin_n0_K}
\end{figure*}

Figure~\ref{fig:isospin_n0_K} shows the asymmetry dependence of the saturation density, $n_0(\diso)$ [panel~(a)], and incompressibility, $K(\diso)$ [panel~(b)]. 
The green markers and shaded bands denote the GP posterior median and the corresponding 68\% and 95\% credibility intervals, respectively.
For comparison, the dashed and dotted lines depict our corresponding results using the ``Meta-model'' (dotted line) and ``Smooth Spline'' method (dashed line). 
For clarity, Fig.~\ref{fig:isospin_n0_K} shows these results only for three representative interactions: the interaction with the lowest SRG resolution scale, ``$(1.8/2.0)$''; the one with the larger 3N cutoff, ``$(2.0/2.5)$''; and the one with different $\pi N$ coupling constants, ``$(2.0/2.0)^*$''~\cite{Hebeler:2010xb}.
These results are consistent with our GP-based constraints at the 68\% credibility level, except for $n_0(\diso)$ obtained from the ``$(2.0/2.0)^*$'' interaction, which nevertheless remains within the 95\% credibility band. 
Naively, one would expect roughly one third of the results obtained from individual EOS calculations to lie outside the 68\% credibility bands; 
however, the limited sample size and the correlations across $\diso$ make this expectation difficult to assess quantitatively.

A few more comments on Fig.~\ref{fig:isospin_n0_K}.
Panel~(a) shows that asymmetric matter exhibits nuclear saturation for $\diso \lesssim 0.7$, consistent with visual inspection of Fig.~\ref{fig:anm_mbpt}.
At the largest shown $\diso = 0.7$, the saturation density has decreased to $n_0(\diso) \approx 0.085 \pm 0.019 \fmiq$.
Panel~(b) confirms that $K(\diso)$ exhibits an approximately quadratic dependence on $\diso$ over the range relevant for medium-mass to heavy nuclei, as expected from the expansion in Eq.~\eqref{eq:K-delta-exp}. 
Furthermore, the negative curvature observed in Fig.~\ref{fig:isospin_n0_K}(b) implies $K_\tau < 0$.
To validate our constraint on $K_\tau$ summarized in Fig.~\ref{fig:corner_emp_eos_params}, we extract $K_\tau$ by fitting parabolas to our samples for $K(\diso \leqslant 0.3)$, resulting in
$K_\tau = -410_{-365}^{+253}\MeV$.
This constraint is statistically consistent with the values obtained from Eq.~\eqref{eq:K_tau}.
Both methods are subject to significant uncertainties.

\subsection{Neutron star EOS}
\label{sec:constaints_ns_eos}

\begin{figure*}[tb]
    \includegraphics[]{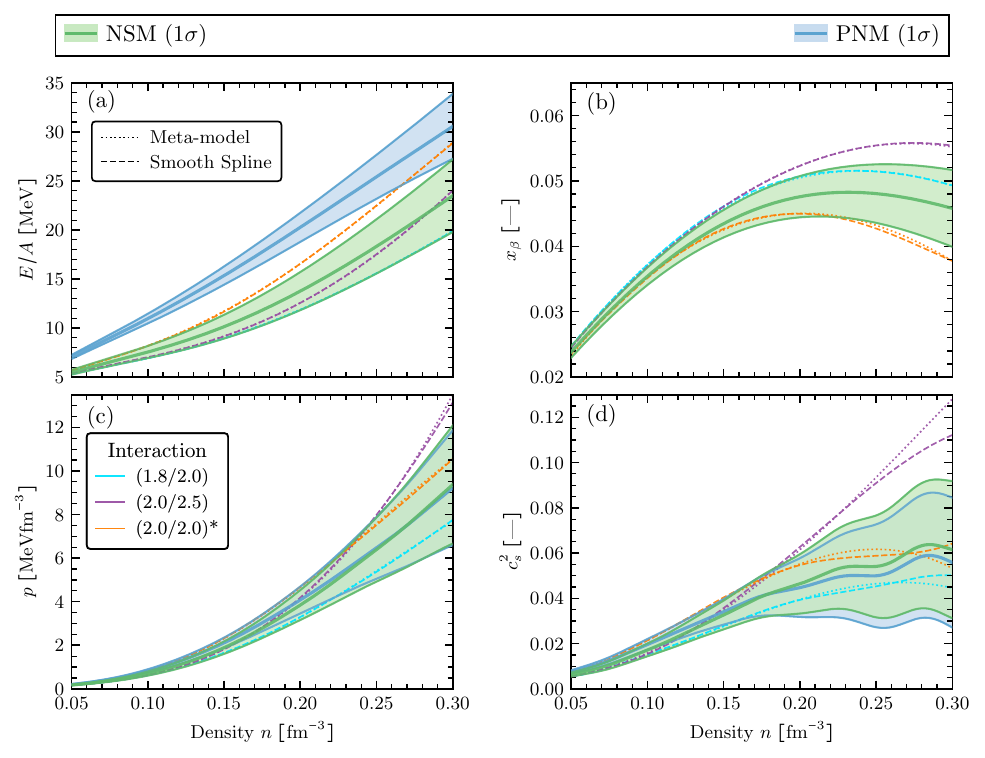} 
    \caption{%
    Constraints on charge-neutral, $\beta$-equilibrated NSM ($\diso=\delta_\beta$; green bands), with PNM constraints ($\diso=1$; blue bands) serving only as a reference here.
    The four panels show the 
    energy per particle $E/A$ [panel~(a)], 
    proton fraction $x_\beta$ [panel~(b)],
    pressure $p$ [panel~(c)], and
    sound speed squared $c_s^2$ [panel~(d)]
    as a function of the nucleon density. 
    The green solid lines and regions depict the predicted mean and $1\sigma$ uncertainty interval from the GP analysis. 
    For comparison, the corresponding results in NSM using the ``Meta-model'' and the ``Smooth Spline'' method are depicted by the dotted and dashed lines, respectively, for the three representative EOS calculations.%
    }
    \label{fig:ns_eos}
\end{figure*}

We constrain the EOS of neutron-star matter~(NSM) at $n \lesssim 2 n_0$, assuming only nucleon and electron degrees of freedom.
At a given $n$, the isospin asymmetry associated with charge-neutral, $\beta$-equilibrated matter, $\delta_\beta(n)$, is determined by
\begin{multline} \label{eq:beta_equi}
    f_{\beta} \left( n, \diso \right) = 0 \\ = \left[- 2 \frac{\partial}{\partial \diso} \frac{E}{A} + \mu_e(\diso, n) - \Delta m \right]_{\diso = \delta_\beta (n)} \,,
\end{multline}
where $\mu_e(\diso, n) = \sqrt[3]{\frac{3\pi^2}{2}(1 - \diso) n}$ is the chemical potential of ultra-relativistic electrons and $\Delta m = m_n-m_p \approx 1.29 \MeV$ is the nucleon mass difference.
Since $x$ and $\diso$ are directly related, we use them interchangeably and denote the corresponding proton fraction in $\beta$-equilibrium as $x_{\beta} (n) = \left( 1 - \delta_{\beta} (n)\right) / 2$.

In addition to the NSM energy per particle, $E(\delta_\beta(n),n)/A$, we constrain its pressure, chemical potential, and speed of sound squared by evaluating the following expressions,
\begin{align}
    p (n) &= n^2 \left. \frac{\partial}{\partial n} \frac{E}{A} \left(\diso,n\right) \right|_{\diso = \delta_{\beta} \left(n\right)} + \frac{1}{4} \mu_e \left( n, \delta_{\beta} \right) x_{\beta} n\,,\label{eq:pressure}\\
    \mu_c (n) &= \left( 1 + n \frac{\partial}{\partial n} \right) \frac{E}{A} \left(\diso,n \right) \bigg|_{\diso = \delta_{\beta} \left(n\right)} \nonumber\\
    &\quad + m + \frac{\Delta m}{2} \delta_{\beta} + \mu_e(\delta_{\beta},n) x_{\beta} \,,\label{eq:chemical_potential} \\
    c_s^2 (n) &= \dv{p(n)}{n} \left[ \mu_c (n) \right]^{-1} \,, \label{eq:cs2_option2}
\end{align}
with the average nucleon mass $m = (m_n+m_p)/2 \approx 938.9 \MeV$.
However, $E(\delta_\beta(n),n)/A$ is in general not a GP anymore, and thus constraining derived quantities in NSM requires sampling, as described in the following.  
Similar to Sec.~\ref{sec:constaints_eos_params}, we jointly sample from the trained GP the energy per particle and the derivatives needed to evaluate these NSM observables.
The samples are evaluated at $n \in [0.05, 0.32] \fmiq$ (28 points) and $\diso \in [0.8, 1.0]$ (21 points) to focus on neutron-rich matter at densities relevant for the outer core.
Specifically, at each density, we first determine $\delta_\beta(n)$ by solving Eq.~\eqref{eq:beta_equi}. 
We bracket the root using the nearest sign change on the asymmetry grid and construct a local cubic spline to determine the root more accurately. 
The derivatives, which are jointly sampled from the GP, are then interpolated using cubic splines, resulting in a data set comprised of the samples $\left\{ n, \delta_{\beta}, \left. \frac{E}{A} \right|_{\diso = \delta_\beta (n)}, \left. \frac{\partial}{\partial n} \frac{E}{A} \right|_{\diso = \delta_\beta (n)} \right \}$.
This dataset is sufficient to determine the energy per particle, pressure, and the composition of NSM.
To also evaluate the sound speed squared~\eqref{eq:cs2_option2}, we furthermore calibrate GPs to the subsets $ \left\{ n, \delta_{\beta} \right \} $, $ \left\{ n, \left. \frac{E}{A} \right|_{\diso = \delta_\beta (n)} \right \} $, and $ \left\{ n, \left. \frac{\partial}{\partial n} \frac{E}{A}\right|_{\diso = \delta_\beta (n)} \right \} $, respectively, using the statistical model described in Sec.~\ref{sec:discrepancy_model}.
For the kernel matrix $\vb{K}_{\delta, tt}$, we use the empirical kernel~\eqref{eq:kdelta_empirical} with the $M = 7$ dominant eigenmodes obtained by diagonalizing the empirical covariance matrix of the samples (cf.\ Sec.~\ref{sec:deviation_kernel}).
Combining these GPs with the propagated covariances derived in Appendix~\ref{app:UP_nsEOS}, e.g., for $\mu_c$, allows us to constrain $c_s^2(n)$.

Figure~\ref{fig:ns_eos} shows the resulting energy per particle $E(\delta_\beta(n),n)/A$ [panel~(a)], proton fraction $x_\beta(n)$ [panel~(b)], pressure $p(n)$ [panel~(c)], and sound speed squared $c_s^2(n)$ [panel~(d)] as a function of the density in NSM (green bands) and PNM (blue bands) up to $n = 0.30 \fmiq$.
The uncertainty bands encompass the $1\sigma$ credibility regions.
In addition to these GP-based bands, we show the corresponding NSM constraints from the ``Meta-Model'' (dotted lines) and the ``Smooth Spline'' method (dashed lines) in Fig.~\ref{fig:ns_eos}, shown again only for the three representative interactions for clarity.

As shown in Fig.~\ref{fig:ns_eos}(a), we obtain $E/A = 10.77 \pm 1.35 \MeV$ at $n = 0.16 \fmiq$ and $E/A = 23.49 \pm 3.70 \MeV$ at $n = 0.30 \fmiq$, which is, within the 1$\sigma$ credibility regions, less than the PNM energy per particle (blue band) across the entire density.
Consistent with the constraints derived in Ref.~\cite{Drischler:2026vdm}, we find that NSM does not exceed $x_\beta \lesssim 0.055$ up to $\approx 2 n_0$ and decreases at $n\gtrsim 1.5 n_0$, as shown in panel~(b). 
That is, we obtain $x_\beta = 0.045 \pm 0.002$ at $n = 0.16 \fmiq$ and $x_\beta = 0.046 \pm 0.006$ at $n = 0.30 \fmiq$.
In contrast to the energies per particle, the pressures in NSM and PNM are comparable in both their inferred mean values and uncertainties. 
Only small differences appear near $n_0$, where the PNM EOS is somewhat stiffer, and near $2n_0$, where the NSM EOS is somewhat stiffer.
This behavior is qualitatively similar to the findings in Ref.~\cite[see Figure~1(b)]{Drischler:2020fvz}.
For NSM, we obtain $p = 2.22 \pm 0.34\MeV\fmiq$ at $n = 0.16\fmiq$ and $p = 9.38 \pm 2.71\MeV\fmiq$ at $n = 0.30\fmiq$. 
For PNM, the corresponding values are $p = 2.45 \pm 0.29\MeV\fmiq$ and $9.23 \pm 2.63\MeV\fmiq$, respectively.
Finally, for the sound speed we obtain the constraints $c_s^2 = 0.036 \pm 0.007$ at $n = 0.16\fmiq$ and $0.061 \pm 0.030$ at $n = 0.30\fmiq$.
We emphasize that, as described in Sec.~\ref{sec:micro_data}, the uncertainty bands in Fig.~\ref{fig:ns_eos} do not explicitly include EFT truncation errors~\cite{Drischler:2020hwi,Drischler:2020yad};
incorporating these errors should be revisited in future work. 
Accordingly, the uncertainty bands should be interpreted as lower bounds on the total theoretical uncertainties, especially at $n \gtrsim 1.5n_0$.

As shown in Fig.~\ref{fig:ns_eos}, the ``Meta-Model'' and the ``Smooth Spline'' method are generally consistent with each other.
Only the sound speed square at high densities [panel~(d)] shows a noticeable sensitivity to the specific extraction method due to the numerical noise in the input data. 
For example, the two predictions for $c_s^2(n = 0.30\,\fmiq)$ based on the ``(2.0/2.5)'' interaction (purple lines) differ slightly compared with the overall spread of the GP-free results, $c_s^2(n = 0.30\,\fmiq) \approx 0.05 - 0.12$.
Nonetheless, the results of both GP-free methods are within the $\approx 2\sigma$ confidence region of our GP-based constraint.
The effects of numerical noise are also manifested in the GP-based constraints for $c_s^2$, where the uncertainty bands exhibit some fluctuations at $n \gtrsim n_0$.
Overall, the agreement of these methods further supports the robustness of our constraints derived from our statistical model and the noisy input data.

\subsection{Crust-core transition}
\label{sec:constaints_cc_trans}

We constrain the crust-core transition density, $\ncc$, below which NSM is unstable against density and composition fluctuations and transitions to an inhomogeneous clustered phase~\cite{ravenhall1983structure,chamel2008physics}.\footnote{%
For a selection of work incorporating constraints from chiral EFT on the crust-core transition and the implications for neutron star properties, we refer the reader to Refs.~\cite{Gottling:2025ohe,Grams:2021qpj,Carreau:2019zdy,Lim:2017luh,Tews:2016ofv,Hebeler:2013nza}.%
}
Its location, e.g., directly impacts the fractional moment of inertia of the crust, which is key to modeling pulsar glitches~\cite{link1999pulsar}. 
Here, we adopt a formulation equivalent to the thermodynamical method~\cite{lattimer2007neutron}, but more concise since it directly tests the local convexity of the energy in the $x$--$n$ [or $\diso$--$n$] plane. 
It accounts for instabilities arising from arbitrary infinitesimal coupled perturbations $(\delta x, \delta n)$. 
Specifically, following Refs.~\cite{Gibbs1878On,kubis2007nuclear,Kunjipurayil:2026pna}, we study the determinant of the Hessian matrix $H_E$ of the energy density $nE(\diso,n)/A$ with respect to $(n\diso,n)$:\footnote{%
Note that $\det[nE(\diso,n)/A]$ with respect to $(n\diso,n)$ is proportional to $\det[nE(n_p,n_n)/A]$ with respect to $(n_p,n_n)$.%
}
\begin{equation} \label{eq:hessian_det}
\det[H_E] = \left(\frac{2}{n}\frac{\partial E}{\partial n}+\frac{\partial^2 E}{\partial n^2}\right) \frac{\partial^2 E}{\partial \diso^2} - \left(\frac{\partial^2 E}{\partial n \partial \diso}\right)^2 \,,
\end{equation}
where we omitted ``$(\diso,n)/A$'' for brevity. 
Uniform nuclear matter is stable when $\det[H_E] \geqslant 0$ and unstable when $\det[H_E] < 0$; $\det[H_E] = 0$ marks the crust-core transition. 
The electron energy can be neglected when evaluating the Hessian determinant~\eqref{eq:hessian_det}, assuming the electrons remain in $\beta$-equilibrium with the nucleons during these perturbations~\cite{kubis2007nuclear}. 
The energy involving density-gradient and Coulomb terms may be included, as in the dynamical method~\cite{ducoin2007isospin,Xu:2008vz,ducoin2011core,Xu:2009vi,ducoin2010nuclear}. 
These terms generally shift the instability toward lower densities, with the resulting transition density typically lower than the thermodynamical estimate, roughly at the $10\%$ level~\cite{Xu:2008vz,ducoin2011core}. 
In this initial application of \gpdiff, however, such finite-size contributions are not incorporated. 
Their impact is expected to be subleading compared with the bulk EOS dependence, particularly the density dependence of the symmetry energy~\cite{Xu:2009vi,ducoin2010nuclear}. 
Hence, we consider only the thermodynamic estimate here.

\begin{figure}[tb]
\includegraphics[width=\linewidth]{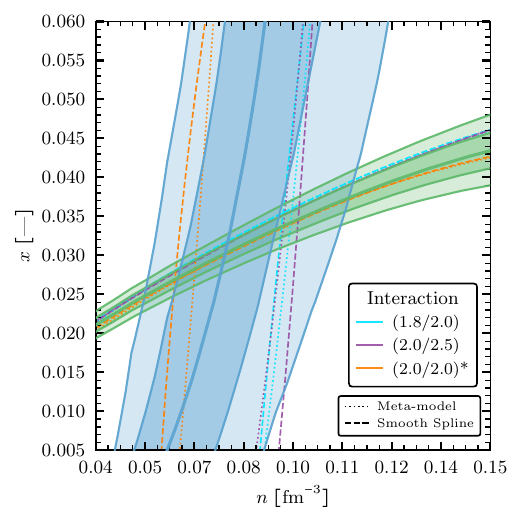}
    \caption{%
    Composition of NSM and instability boundary as a function of the density.
    The green solid line and uncertainty bands show the proton fraction in charge-neutral, $\beta$-equilibrated NSM ($x_\beta$; determined from Eq.~\eqref{eq:beta_equi}), while the blue solid line and bands indicate our estimates of the instability condition ($\ncc$; see Eq.~\eqref{eq:hessian_det}).
    Solid lines represent the median, and dark-shaded (light-shaded) bands depict the 68\% (95\%) confidence regions.
    The intersection of these two uncertainty bands determines the crust-core transition density $\ncc$ and associated proton fraction $\xcc$ in NSM [see Fig.~\ref{fig:corner_crust_core} for the results].
    For comparison, the other lines indicate the corresponding results based on the three representative EOS calculations extracted using the ``Meta-model'' (dotted lines) and ``Smooth Spline'' method (dashed lines).%
    }
    \label{fig:crust-core}
\end{figure}

\begin{figure*}[p]
\includegraphics[width=\linewidth]{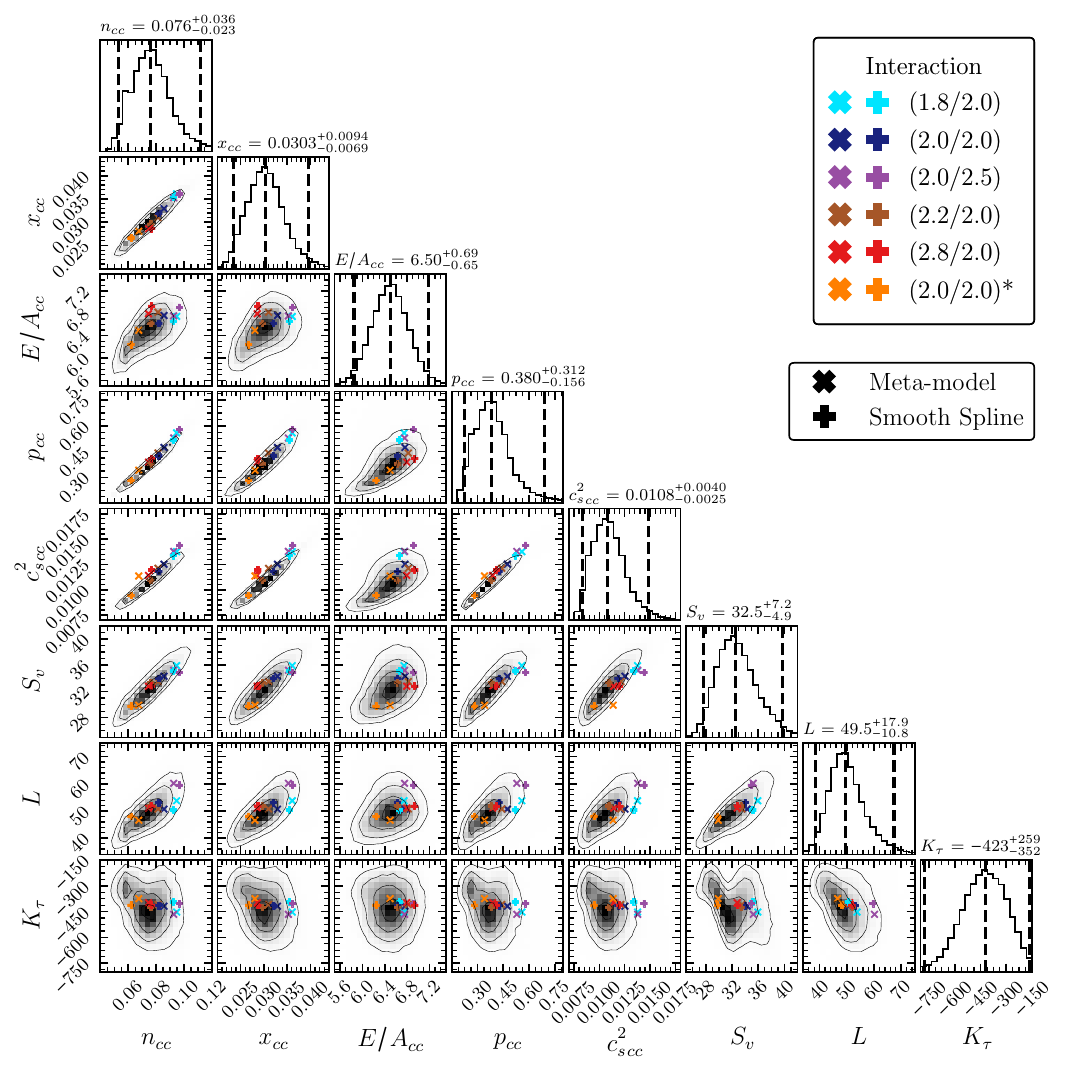}
    \caption{%
    Posterior distributions of the crust-core transition properties and low-density EOS parameters. 
    The crust-core transition is characterized by the transition density $\ncc\ [\text{fm}^{-3}]$ and associated proton fraction $\xcc\ [\textrm{---}]$, energy per particle $E/A_{cc}\ [\MeV]$, pressure $p_{cc}\ [\MeV \fmiq]$, and sound speed squared ${c_s^2}_{cc}\ [\textrm{---}]$. 
    The other EOS parameters are the symmetry energy $S_v\ [\MeV]$, slope parameter $L\ [\MeV]$, and isospin incompressibility parameter $K_\tau\ [\MeV]$ evaluated at the saturation density, corresponding to the results shown in Fig.~\ref{fig:corner_emp_eos_params} up to statistical fluctuations (because of the smaller sampling size). 
    Diagonal panels show marginalized posterior distributions, with titles reporting the median and central 95\% credibility interval. 
    The dashed lines indicate the medians and the bounds of the 95\% credibility intervals, and the contours correspond to the approximate $0.5\sigma$, $1\sigma$, $1.5\sigma$, and $2\sigma$ confidence regions. 
    Symbols indicate values extracted using the ``Meta-model''~($\bm{\times}$) and the ``Smooth Spline''~($\bm{+}$) method based on the six individual EOS calculations.%
    }
    \label{fig:corner_crust_core}
\end{figure*}

We follow a sampling approach to trace the instability condition in the $\diso$--$n$ plane and then to constrain the corresponding crust-core (cc) transition in NSM.
Using \gpdiff, we jointly sample the trained GP and all derivatives necessary to evaluate Eq.~\eqref{eq:hessian_det} on a suitable $(\diso,n)$-grid that covers the region in which we expect the crust-core transition to occur:
$n \in [0.01, 0.17] \fmiq$ (17 points) and $\diso \in [0.8, 0.99]$ (11 points). 
For each joint GP sample, we use bivariate interpolation to locate the zero crossing of $\det[H_E]$ on a dense grid and keep only those samples that exhibit a negative-to-positive zero crossing across the considered density range consistent with the physical boundary of the instability.
At the same time, we determine $\delta_\beta(n)$ by finding the zero crossing of $f_{\beta}$ [see Eq.~\eqref{eq:beta_equi}], as before in Sec.~\ref{sec:constaints_ns_eos}.
Samples for which no zero crossing is found are discarded. 
The crust-core transition in NSM, $(\ncc,\dcc)$, is then determined by the intersection of the $\beta$-equilibrium region and the stability boundary [see Fig.~\ref{fig:crust-core}].
Furthermore, to study correlations between $(\ncc,\dcc)$ and the EOS parameters discussed in Sec.~\ref{sec:constaints_eos_params} [see Fig.~\ref{fig:corner_crust_core}], we extend this approach to include joint samples of the SNM EOS. 
These additional correlated samples are used to determine the saturation point and, subsequently, the associated EOS parameters.

Figure~\ref{fig:crust-core} depicts the constraints on the instability boundary (blue bands) and the proton fraction of NSM (green bands) as a function of the density, resulting from this sampling procedure.
The intersection of these two constraints determines the crust-core transition density $\ncc$ and associated proton fraction $\xcc$ in NSM, which we will discuss below [see also Fig.~\ref{fig:corner_crust_core}].
Dark and light shaded bands indicate the 68\% and 95\% confidence regions, respectively.
Figure~\ref{fig:crust-core} shows that the $\beta$-equilibrium region (green bands) forms a relatively narrow and smoothly increasing band in the $x$--$n$ plane.
In contrast, the stability boundary region (blue bands) exhibits a much larger spread, particularly in the density due to the stronger density dependence.
As a consequence, the uncertainty in the crust-core transition density is mainly driven by the uncertainty in the stability boundary.
This conclusion is also supported by the results obtained with the two GP-free methods (dashed and dotted lines), although their predictions for the stability boundary are somewhat sensitive to numerical noise in the input data.

Figure~\ref{fig:corner_crust_core} shows the inferred joint posterior distributions of the crust-core transition properties and EOS parameters studied in Sec.~\ref{sec:constaints_eos_params}.
Specifically, in addition to $n_{cc}$ and $x_{cc}$, the crust-core transition is characterized by the associated energy per particle $E/A_{cc}$, pressure $p_{cc}$, and sound speed squared ${c_s^2}_{cc}$. 
As before, symbols indicate values extracted using the ``Meta-model'' and the ``Smooth Spline'' method based on the six individual EOS calculations.
The diagonal panels show the marginalized distributions, and their titles report our constraints on these parameters at the 95\% credibility level. 
In particular, we determine $n_{cc} = 0.076_{-0.023}^{+0.036} \fmiq$, which is consistent with the range recently obtained from a chiral EFT truncation error analysis, $n_{cc} \in [0.062, 0.088] \fmiq$~\cite{Gottling:2025ohe}
and from a Bayesian analysis of a compressible liquid-drop model (CLDM), $n_{cc}= 0.072\pm 0.022 \fmiq$ ($2\sigma$-level)~\cite{Carreau2019}.
The latter study also reported a compatible yet somewhat lower $p_{cc} = 0.339 \pm 0.230 \MeV \fmiq$~\cite{Carreau2019} than we obtain, $p_{cc} = 0.380_{-0.156}^{+0.312} \MeV \fmiq$ (both at the 95\% level). 
Furthermore, using a CLDM with additional constraints from the MBPT calculations in Ref.~\cite{Drischler:2015eba} for the chiral interactions ``(2.0/2.0)'' and ``(2.0/2.5)'' employed in this work, Ref.~\cite{Grams:2021qpj} found $n_{cc} = 0.084 - 0.093 \fmiq$, which is also consistent with our estimate within the uncertainties.

The results obtained with the GP-free methods are mostly within the 95\% credibility regions of the inferred GP constraints, indicating that the GP-based uncertainty bands encompass the interaction-dependent variation in the extracted parameters within the estimated uncertainties. 
On the other hand, our nonparametric GP constraints revealed a lack of flexibility of the parametric ``Meta-model'' in constraining the crust-core transition properties. 
Specifically, we found that the ``Meta-model'' without the fourth-order terms in Eq.~\eqref{eq:e-a-semiana}, corresponding to the functional forms used in Refs.~\cite{Drischler:2017wtt,Drischler:2026vdm}, can precisely reproduce the results shown in Figs.~\ref{fig:corner_emp_eos_params}, \ref{fig:isospin_n0_K}, and~\ref{fig:ns_eos}. 
However, it leads to significant deviations from the GP and ``Smooth Spline'' approaches when the Hessian determinant in Eq.~\eqref{eq:hessian_det} is evaluated to determine the crust-core transition properties.
These systematic deviations are remedied by the inclusion of the fourth-order terms in Eq.~\eqref{eq:e-a-semiana}, as shown in Fig.~\ref{fig:corner_crust_core}. 
These comparisons between different methods provide valuable consistency checks of the inferred constraints and underscore the importance of uncertainty estimates, as provided by \gpdiff.


Previous studies have found that $n_{cc}$ generally decreases with increasing $L$ when individual nuclear models or restricted model families are considered~\cite{Xu:2008vz}. 
However, more recent systematic investigations have shown that $n_{cc}$ and $p_{cc}$ cannot be characterized by merely one EOS parameter such as $L$ and $K_{\text{sym}}$ (or the related $K_\tau)$, but instead depend on correlations among several EOS parameters~\cite{ducoin2011core,li2020curvature}. 
Because our GP framework quantifies and propagates the correlations in the underlying nuclear calculations, it enables us to determine how crust-core transition properties near $n_0/2$ correlate with EOS parameters evaluated at $n_0$.

Figure~\ref{fig:corner_crust_core} provides insights into these correlations. 
For example, we find a strong correlation between $n_{cc}$ and $x_{cc}$ because $x_\beta(n)$ is a monotonically increasing function in the vicinity of the crust-core transition, as shown in Fig.~\ref{fig:crust-core}, and both $n_{cc}$ and $x_{cc}$ are correlated with $S_v$ and $L$. 
Interestingly, the observed correlation between $n_{cc}$ and $L$ deviates from the anti-correlation often reported in phenomenological studies based on energy density functionals~\cite{Xu:2008vz,ducoin2010nuclear}, and $n_{cc}$ and $K_\tau$ are uncorrelated.
We also observe strong correlations between $p_{cc}$, $n_{cc}$, ${c_s^2}_{cc}$, and $E/A_{cc}$ since the neutron-rich matter EOS is well constrained at these low densities, as can be seen in Fig.~\ref{fig:anm_mbpt}.

\section{Summary and outlook}
\label{sec:summary_outlook}

\begin{table}[tb]
\renewcommand{\arraystretch}{1.6}
\caption{%
Summary of our constraints on the low-density EOS parameters.
The table shows the 95\% confidence intervals of the marginalized distribution functions. 
We emphasize that, as shown in Fig.~\ref{fig:corner_emp_eos_params}, these parameters are correlated. 
The acronym ``CCT'' stands for crust-core transition.%
}
\label{tab:final_results}
\begin{ruledtabular}
\begin{tabular}{llll}
Symbol     & Parameter               & Constraint                   & Unit         \\ \colrule 
$n_0$      &  Saturation density & $0.169_{-0.037}^{+0.061}$    & $\fmiq$      \\
$E_0/A$    & Saturation energy           & $-15.5_{-4.6}^{+3.0}$     & $\MeV$       \\
$K$        & Incompressibility           & $221_{-67}^{+114}$           & $\MeV$       \\
$S_v$      & Symmetry energy             & $32.6_{-4.9}^{+7.7}$         & $\MeV$       \\
$L$        & Slope parameter             & $49.5_{-10.9}^{+18.5}$       & $\MeV$       \\
$K_\tau$   & $K$'s isospin dependence    & $-423_{-374}^{+263}$         & $\MeV$       \\
\hline
$n_{cc}$   & CCT density         & $0.076_{-0.023}^{+0.036}$ & $\fmiq$          \\
$x_{cc}$   & CCT proton fraction         & $0.030_{-0.007}^{+0.009}$ & ---          \\
$E/A_{cc}$ & CCT energy per particle     & $6.5_{-0.7}^{+0.7}$       & $\MeV$       \\
$p_{cc}$   & CTT pressure                & $0.38_{-0.16}^{+0.31}$    & $\MeV \fmiq$
\end{tabular}
\end{ruledtabular}
\end{table}

We have presented microscopic constraints on the low-density nuclear EOS at zero temperature with quantified correlated uncertainties based on a novel GP framework, named \gpdiff. 
Specifically, we have studied the saturation point $(n_0,E_0/A)$ in SNM, the incompressibility $K$ and its leading isospin dependence $K_\tau$, the symmetry energy $S_v$ and its slope parameter $L$ [see Sec.~\ref{sec:constaints_eos_params}], as well as properties of charge-neutral, $\beta$-equilibrated NSM, including its EOS [see Sec.~\ref{sec:constaints_ns_eos}] and crust-core transition density [see Sec.~\ref{sec:constaints_cc_trans}].
Table~\ref{tab:final_results} summarizes the main results of our analysis at the 95\% credibility level.
In particular, we look forward to the forthcoming constraints on $K_\tau$, which is only weakly determined in our analysis, obtained from the FRIB experiment ``The Isoscalar Giant Monopole Resonance in \isotope[132]{Sn}: Implications on the Nuclear Incompressibility''~\cite{Brown:2024rml}.
These results will provide valuable benchmarks for microscopic nuclear interactions. 

To this end, we have developed \gpdiff, a \texttt{JAX}-based Python package for multivariate GP regression with automatic differentiation and support for user-defined kernels, including input-dependent kernels.
Once trained on, e.g., microscopic calculations of the energy per particle, \texttt{GPDiff} enables joint predictions of the EOS and derivatives of arbitrary order with respect to the input variables, including mixed partial derivatives; 
thereby propagating correlated uncertainties from these many-body calculations, which may contain numerical noise in addition to other theoretical uncertainties, to derived observables, such as the pressure and speed of sound.
In that regard, \gpdiff provides an improved, more flexible Python implementation of GP regression with arbitrary derivative predictions, particularly for EOS applications, than the earlier package \texttt{gptools}~\cite{Chilenski_2015_gptools}, which, to our knowledge, is no longer actively maintained.

As a first application, we trained \gpdiff on recent high-order MBPT calculations of asymmetric nuclear matter~\cite{Drischler:2026vdm} based on the six Hebeler~et al.\ interactions~\cite{Hebeler:2010xb}. 
To analyze the corresponding set of distinct EOS calculations, we have developed a statistical model [see Sec.~\ref{sec:discrepancy_model}] that separates the common mean EOS shared by these six EOS from their interaction-dependent deviations and numerical noise in the MBPT calculations. 
This model allowed us to construct predictive uncertainty bands that account for correlations across density and isospin asymmetry, and uncertainties due to the limited sample size of only six individual EOS calculations.
Future work may apply \gpdiff in the opposite, data-rich limit by leveraging fast and accurate emulators for many-body calculations~\cite{Duguet:2023wuh,Jiang:2022oba,Jiang:2022tzf}, including parametric matrix models~\cite{Cook:2024toj,Somasundaram:2024zse,Armstrong:2025tza,Curry:2025pna,Cook:2026yhj}, to propagate uncertainties in the low-energy couplings, together with EFT truncation errors, to the nuclear EOS.

To model the mean EOS, we have explored input-dependent CS kernels, which generalize CP kernels~\cite{Semposki:2025etb} to higher-dimensional input spaces, here constructed from multiple RBF kernels.
These kernels led to statistically consistent results compared to those obtained with a single RBF kernel.
This finding suggests that smooth (i.e., infinitely differentiable) kernels with a single length scale in each input dimension can already capture the dominant features of the asymmetric matter EOS, at least at the mean level.
However, when modeling the EOS over a wider density range (e.g., for neutron star applications), the CS kernels may become more important for encoding nonstationarity, variations in smoothness, and related EOS features such as phase transitions.
Interesting stationary kernel choices that could be combined into nonstationary CS kernels for these studies include the Mat\'ern, polynomial, and rational quadratic kernels~\cite{duvenaud_PhD_2014}.

We have compared the inferred GP-based constraints with two independent GP-free methods, a parametric meta-model and a smooth bivariate spline, calibrated separately to the different EOS calculations. 
The observed agreement between these three approaches supports the robustness of the derived low-density EOS parameters and NSM properties in the presence of numerical noise in the many-body calculations, and provides a consistency check of the underlying GP modeling assumptions. 
This comparison also highlighted the importance of uncertainty estimates and nonparametric EOS modeling, both facilitated by \gpdiff. 

Several extensions of our GP framework would be valuable. 
\gpdiff is independent of the idiosyncrasies of the EOS calculations, and so studying results obtained with different many-body frameworks and a wider range of chiral interactions~\cite{Huther:2019ont,Arthuis:2024mnl} would provide insights into many-body uncertainties. 
For example, order-by-order EOS studies based on LENPIC's SMS potentials~\cite{Reinert:2017usi,Epelbaum:2022cyo} and delta-full potentials~\cite{Piarulli:2019cqu,Jiang:2020the,Nosyk:2021pxb} comparing MBPT and nonperturbative calculations would be interesting~\cite{Drischler:2016djf,Marino:2024tfp}.
In this spirit, combining constraints from complementary low-density EOS calculations, including recent QMC results for NSM in the nonuniform region~\cite{Fore:2024exa}, will be important for improving constraints on the crust-core transition.
Furthermore, because \gpdiff already supports multidimensional inputs, its extension to microscopic EOS calculations at finite temperature~\cite{Keller:2020qhx,Keller:2022crb} should be straightforward. 
This extension would enable thermodynamic quantities relevant to supernovae, neutron-star mergers, and heavy-ion collisions to be modeled and extrapolated based on different kernel choices, aimed at assessing the sensitivity of the inferred results to the GP model assumptions encoded in the kernels.

In these studies, EFT truncation errors, which we did not explicitly quantify, should also be incorporated rigorously using the BUQEYE truncation-error model~\cite{Drischler:2020hwi,Drischler:2020yad}.
We expect \gpdiff to be compatible with this error model~\cite{Melendez:2019izc}. 
The uncertainty bands presented here therefore likely underestimate the full theoretical uncertainty at densities above $\approx 1.5n_0$.
In this regard, marginalizing over the posterior distributions of the kernel hyperparameters instead of simply using the corresponding MAP value would also be insightful, though computationally expensive, to account for these additional uncertainties, especially for predictions of EOS derivatives.

Our \gpdiff framework, together with the presented results, will be made publicly available on GitHub~\cite{BUQEYEsoftware}, enabling practitioners to use, adapt, and extend it in their own work. 
With comprehensive documentation and tutorials, we hope that \gpdiff will become a useful and versatile tool for GP-based workflows aimed at constraining the EOS in the multimessenger astronomy era, including Bayesian model mixing~\cite{Semposki:2024vnp,Semposki:2025etb} and nonparametric inference with GP-based priors~\cite{Gorda:2026rzm,Legred:2026zok,Finch:2025bao,Gorda:2025aiu,Legred:2025aar,Ng:2025wdj,Golomb:2024mmt,Essick:2023fso,Legred:2023als,Legred:2021hdx,Landry:2018prl}.

\begin{acknowledgments}
We are grateful to R.~J.~Furnstahl, M.~Grosskopf~\cite{dfdjaxGP}, M.~L.\ Kumamoto, D.~R.\ Phillips, and J.~Piekarewicz for fruitful discussions.
We also thank the N3AS collaboration~\cite{N3AS} for its encouragement and support.
Y.~G.~L.\ and C.~D.\ are supported by the National Science Foundation (NSF) under award PHY-2339043.
J.~K.\ is supported by the U.S.\ Department of Energy, Office of Science, Office of Nuclear Physics, under contracts DE-AC02-06CH11357, by the DOE Early Career Research Program, and under the STREAMLINE~2 Collaboration Award.
T.~Z.\ is supported by N3AS's NSF Award No.\ 2020275. 
This material is in part based upon work supported by the U.S.\ Department of Energy, Office of Science, Office of Nuclear Physics, under the FRIB Theory Alliance award DE-SC0013617.
The following open-source libraries were used to generate the results in this work:
\texttt{corner}~\cite{corner},
\texttt{JAX}~\cite{jax2018github},
\texttt{Jupyter}~\cite{jupyter},
\texttt{matplotlib}~\cite{Hunter:2007},
\texttt{PyNumDiff}~\cite{PyNumDiff},
\texttt{scipy}~\cite{2020SciPy-NMeth}, \texttt{optax}~\cite{deepmind2020jax}, and 
\texttt{numpy}~\cite{harris2020array}.
\end{acknowledgments}

\section*{Data availability}

The data that support the findings of this article will be made openly available~\cite{BUQEYEsoftware}.

\appendix

\section{Existing GP libraries}
\label{app:other_gp_libs}

\begin{table*}[tb]
\renewcommand{\arraystretch}{1.35}
\caption{%
A selection of existing software packages for GP regression and related applications.
For further details, we refer to the respective package documentation, which served as the source for the information given in the ``Description'' column.%
}
\label{tab:software}
\begin{ruledtabular}
\begin{tabular}{llp{0.58\linewidth}}
Package & Engine & Description \\
\colrule
\texttt{George}~\cite{Ambikasaran_georgeGP} 
& C\textsuperscript{++} \& Python 
& Fast and flexible GP regression library focused on efficient marginal-likelihood evaluation, kernel composition, and custom modeling \\

\texttt{GPflow}~\cite{GPflow2017,GPflow2020multioutput} & TensorFlow (Python) & Implements modern GP inference with composable kernels and likelihoods, including exact, sparse, variational, and MCMC-based models \\

\texttt{GPJax}~\cite{Pinder2022gpjax} 
& \texttt{JAX} (Python) 
& Supports GPU acceleration and just-in-time compilation; designed for flexible research prototyping \\

\texttt{gptools}~\cite{Chilenski_2015_gptools} 
& \texttt{NumPy}/\texttt{SciPy} (Python) 
& Regression package with support for derivative observations and predictions, derivative variances, and arbitrary linearly transformed quantities; developed and tested using Python 2.7, which reached end-of-life in 2020; applied by the nuclear theory community, e.g., in  Refs.~\cite{Drischler:2020hwi,Drischler:2020yad,Keller:2022crb}\\

\texttt{GPy}~\cite{gpy2012} 
& \texttt{NumPy} \& Cython (Python) 
& Framework from SheffieldML with support for GP regression, classification, sparse GPs, multi-output models, latent-variable models, and a broad range of kernels \\

\texttt{GPyTorch}~\cite{Gardner_gpytorch} 
& \texttt{PyTorch} (Python) 
& Scalable and modular GP library supporting exact and approximate inference, GPU acceleration, and modern variational methods \\

\texttt{hetGP}~\cite{hetGP} 
& R 
& Heteroskedastic GP regression package that models input-dependent noise and supports sequential design under replication \\

\texttt{MuyGPyS}~\cite{muygps2021} 
& \texttt{NumPy} (Python)
& Toolkit for scalable approximate GP inference using nearest-neighbor sparsification; supports fast training and prediction on large datasets; engines other than \texttt{NumPy} are optional\\

\texttt{scikit-learn}~\cite{scikit-learn} 
& \texttt{NumPy}/\texttt{SciPy} (Python) 
& Standard GP regression and classification implementation; straightforward to use, but non-sparse and somewhat limited in scalability
\end{tabular}
\end{ruledtabular}
\end{table*}

Table~\ref{tab:software} provides a non-exhaustive list of existing GP software libraries, along with a brief description.
Although \gpdiff could, in principle, be added to this listing as a general-purpose GP library, it was specifically designed for applications to EOS modeling, analysis, and inference. 


\section{Uncertainties in the neutron star EOS}
\label{app:UP_nsEOS}

This appendix explains the UQ of neutron star EOS using the trained GPs in Sec.~\ref{sec:constaints_ns_eos}.
Using standard identities for adding normal distributions, the uncertainties of chemical potential can be calculated by performing error propagation as follows:
\begin{equation}
\begin{aligned}
\sigma^2_{\mu_c} \left( n \right) &= \sigma^2_{\left.\frac{E}{A}\right|_{\diso = \delta_{\beta} \left(n\right)}} + n^2 \sigma^2_{\left. \frac{\partial}{\partial n} \frac{E}{A} \right|_{\diso = \delta_{\beta} \left(n\right)}} + V_{\delta_{\beta}}^2 \sigma^2_{\delta_{\beta}} \\
&+ 2 n \, \mathrm{cov} \left( \left. \frac{E}{A} \right|_{\diso = \delta_{\beta} \left(n\right)}, \left. \frac{\partial}{\partial n} \frac{E}{A} \right|_{\diso = \delta_{\beta} \left(n\right)} \right) \\
&+ 2 V_{\delta_{\beta}} \left[ \mathrm{cov}\left( \left. \frac{E}{A} \right|_{\diso = \delta_{\beta} \left(n\right)}, \delta_\beta \right) \right. \\
&\quad\quad\quad\quad+ \left. n \, \mathrm{cov}\left( \left. \frac{\partial}{\partial n} \frac{E}{A} \right|_{\diso = \delta_{\beta} \left(n\right)}, \delta_\beta \right) \right],
\end{aligned}
\label{Eq:neutron_star_traits_error_propagation_chemical_potential}
\end{equation}
where the coefficient is $V_{\delta_{\beta}} = \frac{\Delta m}{2} - \frac{2}{3} \mu_e \left( \bar{\delta}_\beta, n \right)$. 
The uncertainties in the speed of sound squared can be calculated by performing error propagation as follows:
\begin{align}
\sigma^2_{c_s^2} \left( n \right) &= \bar{c_s^2}^2 \left( \frac{\sigma^2_{\frac{d p (n)}{d n}}}{\overline{\frac{d p (n)}{d n}}^2} + \frac{\sigma^2_{\mu_c} \left( n \right)}{\bar{\mu}^2_c \left( n \right)} - 2 \frac{\mathrm{cov}(\frac{d p (n)}{d n}, \mu_c)}{\overline{\frac{d p (n)}{d n}} \cdot \bar{\mu}_c} \right)\,,\label{Eq:neutron_star_traits_error_propagation_speed_of_sound} \\
\sigma^2_{\frac{d p (n)}{d n}} &= 4 n^2 \sigma^2_{\left.\frac{\partial}{\partial n} \frac{E}{A}\right|_{\diso = \delta_{\beta} \left(n\right)}} + n^4 \sigma^2_{\frac{d}{d n} \left. \frac{\partial}{\partial n} \frac{E}{A} \right|_{\diso = \delta_{\beta} \left(n\right)}} \nonumber\\
&+ W_{\delta_{\beta}}^2 \sigma^2_{\delta_{\beta}} + W_{\frac{d \delta_{\beta}}{d n}}^2 \sigma^2_{\frac{d \delta_{\beta}}{d n}} \nonumber\\
&+ 4 n^3 \mathrm{cov} \left( \left. \frac{\partial}{\partial n} \frac{E}{A} \right|_{\diso = \delta_{\beta} \left(n\right)}, \frac{d}{d n} \left.\frac{\partial}{\partial n} \frac{E}{A} \right|_{\diso = \delta_{\beta} \left(n\right)} \right) \nonumber\\
&+ 2 W_{\delta_{\beta}} W_{\frac{d \delta_{\beta}}{d n}} \mathrm{cov} \left( \delta_{\beta}, \frac{d \delta_{\beta}}{d n} \right) \nonumber\\
&+ 4 n W_{\delta_{\beta}} \mathrm{cov} \left( \left. \frac{\partial}{\partial n} \frac{E}{A} \right|_{\diso = \delta_{\beta} \left(n\right)}, \delta_{\beta} \right)\,, \label{Eq:neutron_star_traits_error_propagation_speed_of_sound_numerator}
\end{align}
with the covariance term
\begin{equation}
\begin{aligned}
\mathrm{cov} &\big( \frac{d p (n)}{d n}, \mu_c \big) \\ 
&= 2 n \ \mathrm{cov} \left( \left. \frac{E}{A} \right|_{\diso = \delta_{\beta} \left(n\right)}, \left. \frac{\partial}{\partial n} \frac{E}{A} \right|_{\diso = \delta_{\beta} \left(n\right)} \right) \\
&+ 2 n^2 \ \sigma^2_{\left. \frac{\partial}{\partial n} \frac{E}{A} \right|_{\diso = \delta_{\beta} \left(n\right)}} \\
&+ n^3 \ \mathrm{cov} \left( \left. \frac{\partial}{\partial n} \frac{E}{A} \right|_{\diso = \delta_{\beta} \left(n\right)}, \frac{d}{d n} \left. \frac{\partial}{\partial n} \frac{E}{A}  \right|_{\diso = \delta_{\beta} \left(n\right)} \right) \\
&+ W_{\delta_{\beta}} \ \mathrm{cov} \left( \left. \frac{E}{A}  \right|_{\diso = \delta_{\beta} \left(n\right)}, \delta_\beta \right) \\
&+ n \left( 2 V_{\delta_{\beta}} + W_{\delta_{\beta}} \right) \mathrm{cov} \left( \left. \frac{\partial}{\partial n} \frac{E}{A}  \right|_{\diso = \delta_{\beta} \left(n\right)}, \delta_{\beta} \right) \\
&+ W_{\delta_{\beta}} V_{\delta_{\beta}} \ \sigma^2_{\delta_{\beta}} + W_{\frac{d \delta_{\beta}}{d n}} V_{\delta_{\beta}} \ \mathrm{cov} \left( \delta_{\beta}, \frac{d \delta_{\beta}}{d n} \right).
\end{aligned}
\label{Eq:neutron_star_traits_error_propagation_speed_of_sound_cov_term}
\end{equation}
The coefficients are given by 
\begin{align}
W_{\delta_{\beta}} &= -\frac{2}{9} \mu_e \left( \bar{\delta}_\beta, n \right) + \frac{1}{36} \mu_e \left( \bar{\delta}_\beta, n \right) \frac{n}{\bar{x}_\beta} \overline{{\frac{d \delta_{\beta}}{d n}}} \,,\\
W_{\frac{d \delta_{\beta}}{d n}} &= -\frac{1}{6} \mu_e \left( \bar{\delta}_\beta, n \right) n \,,
\end{align}
where bars indicate mean values.

Both, $\mathrm{cov} \left( \left. \frac{\partial}{\partial n} \frac{E}{A} \right|_{\diso = \delta_{\beta} \left(n\right)}, \left. \frac{d}{d n} \frac{\partial}{\partial n} \frac{E}{A} \right|_{\diso = \delta_{\beta} \left(n\right)} \right) $  and $\mathrm{cov} \left( \delta_{\beta}, \frac{d \delta_{\beta}}{d n} \right)$ are directly obtained from the derivatives of the GP models trained on $\left\{ n, \left. \frac{\partial}{\partial n} \frac{E}{A}\right|_{\diso = \delta_{\beta} \left(n\right)} \right \}$ and $\left\{ n, \delta_{\beta} \right\}$, respectively. 
The three remaining covariance terms in Eq.~\eqref{Eq:neutron_star_traits_error_propagation_chemical_potential}, which also appear in Eqs.~\eqref{Eq:neutron_star_traits_error_propagation_speed_of_sound_numerator} and~\eqref{Eq:neutron_star_traits_error_propagation_speed_of_sound_cov_term} are evaluated at the same $n$ and are extracted by training standard GP interpolants (without deviation kernels since the uncertainties of the covariances themselves are not required), calibrated to the mean of the generated samples.

Note that several cross-covariance terms involve higher-order derivatives across different physical quantities (e.g., between $\left. \frac{E}{A} \right\vert{}_{\diso = \delta_{\beta}}$ and $\frac{d \delta_{\beta}}{d n}$, or between $\delta_{\beta}$ and $\frac{d}{d n} \left. \frac{\partial}{\partial n} \frac{E}{A} \right\vert{}_{\diso = \delta_{\beta}}$) are not included in the equations above. 
This is a direct consequence of the model architecture: 
while covariances among the base variables (e.g., those directly sampled from the GP) and intra-variable derivative covariances (e.g., between a variable and its own derivative) are strictly constrained by the trained GPs, the cross-derivative covariances between independently evaluated models are inaccessible and thus neglected. 

\section{Additional results}
\label{app:add_results}

This appendix contains additional results obtained using a single RBF kernel, supplementing the results presented in Sec.~\ref{sec:results}.
In general, we find qualitatively similar results for these kernels, indicating that the two sets of results are consistent with one another within their uncertainties. 
Hence, the results in this appendix do not change the conclusions made in Sec.~\ref{sec:results}.

\begin{figure*}[p]
    \includegraphics[width=\linewidth]{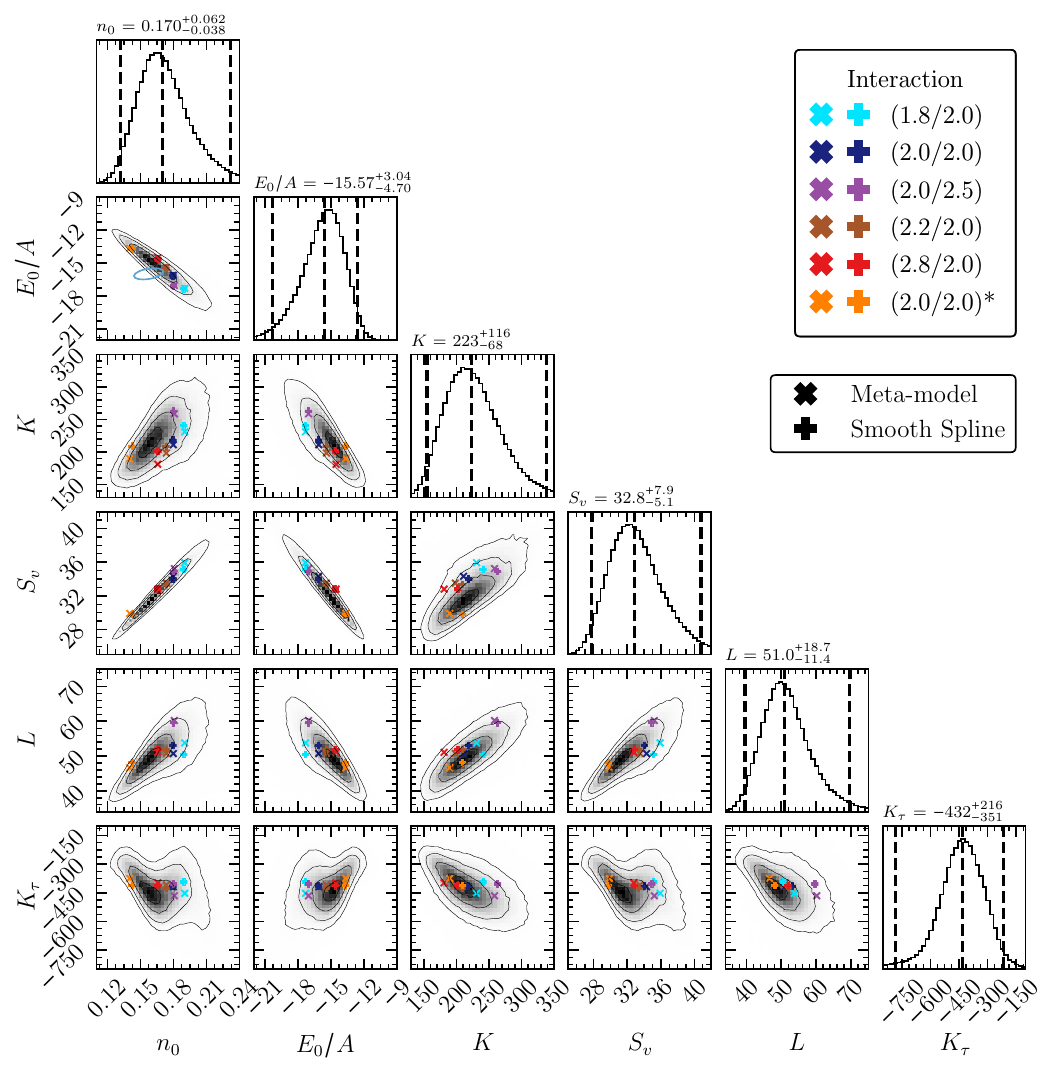}
    \caption{%
    Same as Fig.~\ref{fig:corner_emp_eos_params} but for the single RBF kernel.%
    }
    \label{fig:corner_emp_eos_params_RBF}
\end{figure*}

Figure~\ref{fig:corner_emp_eos_params_RBF} shows the inferred low-density EOS parameters at $n_0$, similar to Fig.~\ref{fig:corner_emp_eos_params}.
Within their uncertainties reported in the titles along the diagonal panels, the inferred constraints on these parameters based on the two kernel choices match. 
The same applies to the agreement between the GP-free extraction methods and our GP-based constraints.
We find, as expected, a strong Coester anti-correlation ($\rho = -0.93$), with the approximate bivariate normal distribution:
\begin{equation} %
   \vb*{\mu}
   \approx \begin{bmatrix}
   0.173 \\ -15.8
   \end{bmatrix} \,, \quad 
   \vb*{\Sigma} \approx \mqty[0.025^2 & -0.215^2 \\ -0.215^2 & 1.98^2] \,.
\end{equation}
Likewise, we find a strong correlation ($\rho = 0.77$) between $S_v$ and L, with the approximate bivariate normal distribution:
%
\begin{equation} %
   \vb*{\mu} 
   \approx \begin{bmatrix}
   33.2 \\ 51.9
   \end{bmatrix} \,, \quad 
   \vb*{\Sigma} \approx \mqty[3.32^2 & 4.40^2 \\ 4.40^2 & 7.61^2] \,.
\end{equation}

\begin{figure*}[tbp]
    \centering
    \includegraphics[width=\linewidth]{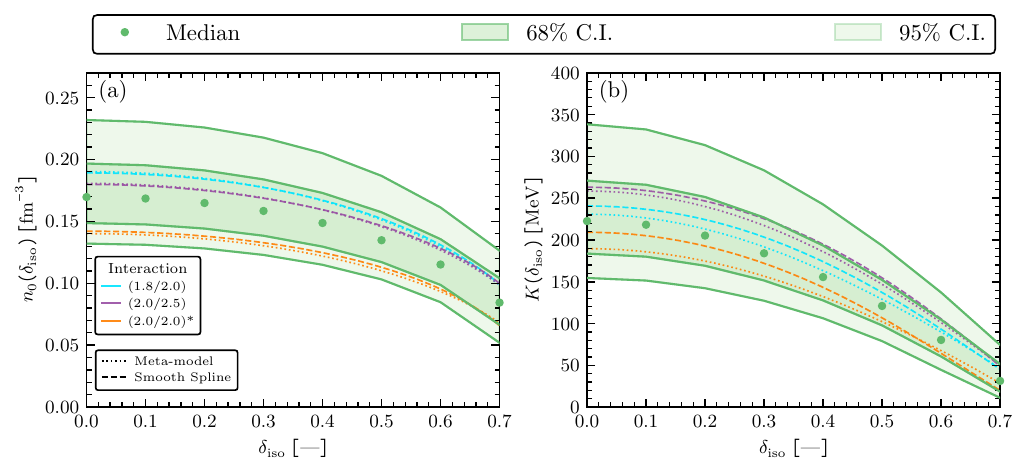}
        \caption{%
        Same as Fig.~\ref{fig:isospin_n0_K} but for the single RBF kernel.%
        }
    \label{fig:isospin_n0_K_RBF}
\end{figure*}

Figure~\ref{fig:isospin_n0_K_RBF} shows the isospin dependence of the saturation point [panel~(a)] and the incompressibility [panel~(b)], similar to Fig.~\ref{fig:isospin_n0_K}.
At the 95\% credibility level, we obtain $n_0(\diso) \approx 0.0844^{+0.0420}_{-0.0325} \fmiq$ at the highest shown $\diso = 0.7$, and $K_\tau = -418_{-346}^{+227}\MeV$ from fitting a parabola [cf.~Eq.~\eqref{eq:K-delta-exp}] to the results in panel~(b) for $\diso \leqslant 0.3$.
These constraints are consistent with our other extractions. 

\begin{figure*}[tb]
    \includegraphics[width=\linewidth]{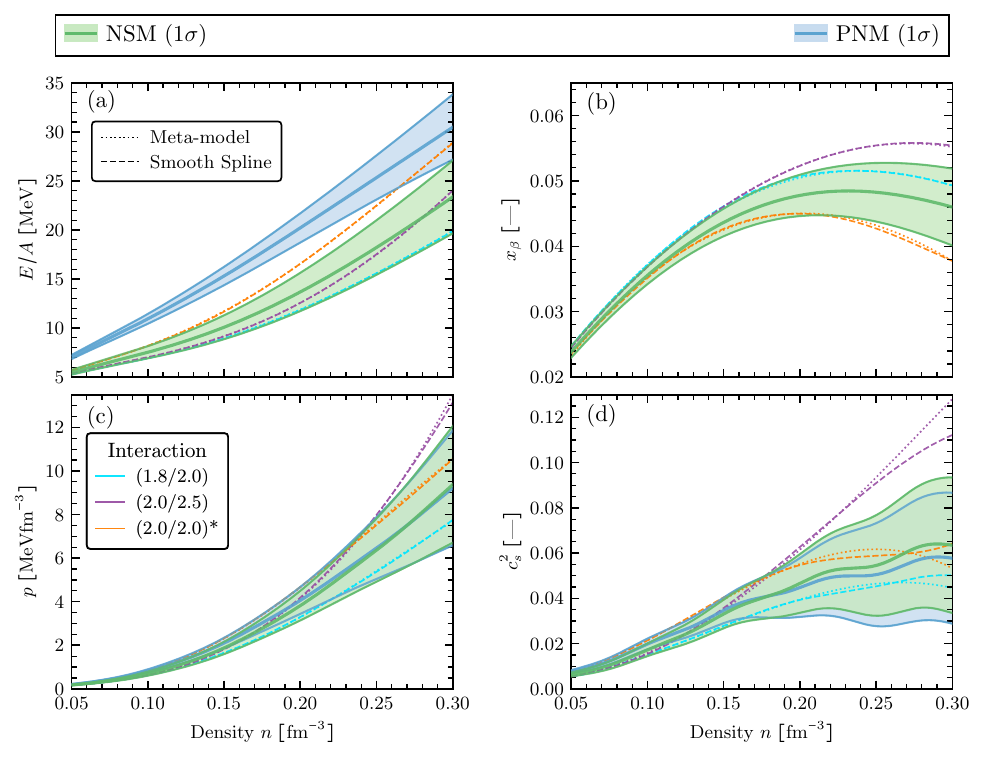} 
    \caption{%
    Same as Fig.~\ref{fig:ns_eos} but for the single RBF kernel.%
    }
    \label{fig:ns_eos_RBF}
\end{figure*}

Figure~\ref{fig:ns_eos_RBF} shows the NSM EOS, similar to Fig.~\ref{fig:ns_eos}.
Matching the results of our extraction in the main text, we find at the 68\% credibility level at the highest density depicted, $n = 0.30 \fmiq$:
$E(n)/A = 23.37 \pm 3.72 \MeV$, 
$p(n) =  9.38 \pm 2.68 \MeV \fmiq$, 
$x_\beta(n) = 0.046 \pm 0.006$, and 
$c_s^2(n) = 0.063 \pm 0.030$.

\begin{figure}[tb]
\includegraphics[width=\linewidth]{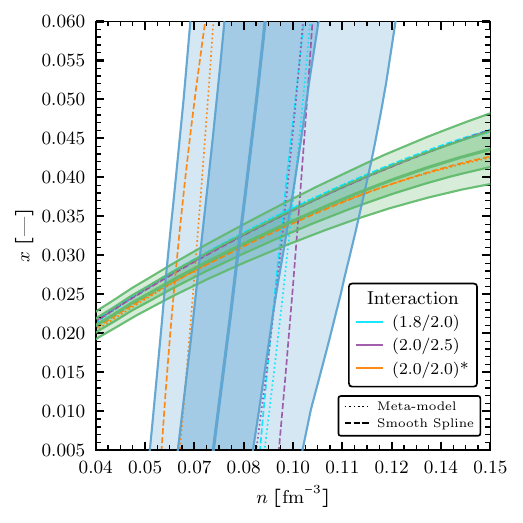}
    \caption{%
    Same as Fig.~\ref{fig:crust-core} but for the single RBF kernel.%
    }
    \label{fig:crust-core_RBF}
\end{figure}

\begin{figure*}[p] 
\includegraphics[width=\linewidth]{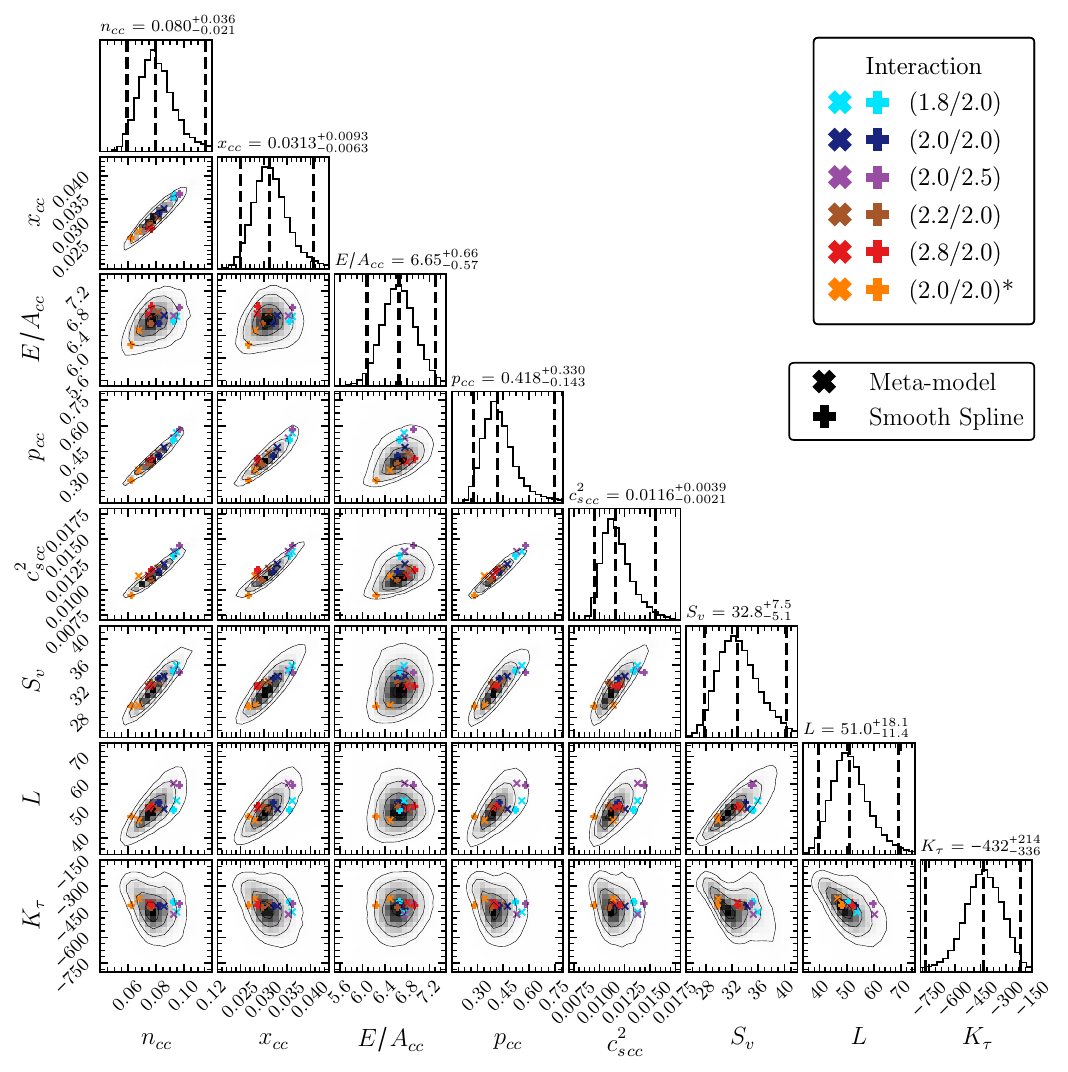}
    \caption{%
    Same as Fig.~\ref{fig:corner_crust_core} but for the single RBF kernel.%
    }
    \label{fig:corner_crust_core_RBF}
\end{figure*}

Figure~\ref{fig:crust-core_RBF} shows the constraints on the crust-core transition boundary (blue bands) and the NSM region (green bands) in the $x$--$n$ plane, similar to Fig.~\ref{fig:crust-core}, while Fig.~\ref{fig:corner_crust_core_RBF} shows the corresponding inferred posterior, analogous to Fig.~\ref{fig:corner_crust_core}.
Our results reported in the titles of the diagonal panels at the 95\% credibility level are consistent within their uncertainties with those reported in Fig.~\ref{fig:corner_crust_core}.

Finally, we also studied different configurations for mixing RBF kernels, including a very expressive CS kernel with 6 RBF kernels, whose centers were located at $n_0/2$ and $3n_0/2$ with $\diso = 0, \pm 1$.
This CS kernel not only distinguishes the EOS features of PNM and SNM, but also separates their behavior at both low and high densities.
The hyperparameters of the kernels at $\diso = \pm 1$ were shared, amounting to 12 free hyperparameters in total.
Within the estimated uncertainties, this kernel configuration led to consistent results compared to the ones presented in this appendix and the main text.
This finding indicates that the CS kernel mixing only two RBF kernels, or even a single RBF kernel, is an adequate compromise between accuracy and model complexity.

\bibliographystyle{apsrev4-2}
\bibliography{bayesian_refs,bib,bib_add}

\end{document}